%% file: main.tex
\documentclass[a4paper,fleqn]{cas-sc}

\usepackage[authoryear,longnamesfirst]{natbib}

\usepackage{graphicx}
\usepackage{float}
\usepackage{array} 
\usepackage{wrapfig}
\usepackage{longtable}
\usepackage{lineno}
\usepackage{framed}
\restylefloat{table}
\usepackage{rotating}
\usepackage{enumitem}
\usepackage{color}

\usepackage{mwe}
\usepackage{subfig}

\usepackage{colortbl}
\usepackage{amsmath}
\usepackage{hyperref}
\hypersetup{colorlinks,allcolors=black}

\usepackage{tabularx}

\usepackage{url}
\usepackage{bbding}
\usepackage{balance}
\usepackage{multirow}
\usepackage{booktabs}
\usepackage{graphicx}
\usepackage[framemethod=tikz]{mdframed}

\usepackage{amssymb}
\usepackage{graphicx}
\usepackage{rotating}

\usepackage{footnote}
\makesavenoteenv{tabular}
\usepackage{tablefootnote}

\def\tsc#1{\csdef{#1}{\textsc{\lowercase{#1}}\xspace}}
\tsc{WGM}
\tsc{QE}

\begin{document}
\let\WriteBookmarks\relax
\def\floatpagepagefraction{1}
\def\textpagefraction{.001}

\shorttitle{Counter-terrorism in Cyber-Physical Spaces: Best Practices and Technologies from the State of the Art}    

\shortauthors{Cascavilla et al.}  

\title [mode = title]{Counter-terrorism in Cyber-Physical Spaces: Best Practices and Technologies from the State of the Art}  



%

\author[1]{Giuseppe Cascavilla}[orcid=0000-0002-0724-3772]
\cormark[1]
\ead{g.cascavilla@tue.nl}

\address[1]{Eindhoven University of Technology, Jheronimus Academy of Data Science, The Netherlands}

\author[1]{Damian A. Tamburri}
\ead{d.a.tamburri@tue.nl}

\author[3]{Francesco Leotta}
\ead{leotta@diag.uniroma1.it}

\author[3]{Massimo Mecella}
\ead{mecella@diag.uniroma1.it}

\author
[2]{WillemJan Van Den Heuvel}
\ead{w.j.a.m.v.d.heuvel@jads.nl}

\address[2]{Tilburg University, Jheronimus Academy of Data Science, The Netherlands}

\address[3]{Sapienza, University of Rome, Italy}















\cortext[1]{Corresponding author}



\begin{abstract}
\noindent \textbf{Context}: The demand for protection and security of physical spaces and urban areas increased with the escalation of terroristic attacks in recent years. We envision with the proposed cyber-physical systems and spaces, a city that would indeed become a smarter urbanistic object, proactively providing alerts and being protective against any threat.\\
\textbf{Objectives}: This survey intend to provide a systematic multivocal literature survey comprised of an updated, comprehensive and timely overview of state of the art in counter-terrorism cyber-physical systems, hence aimed at the protection of cyber-physical spaces. Hence, provide guidelines to law enforcement agencies and practitioners providing a description of technologies and best practices for the protection of public spaces.\\
\textbf{Methods}: We analyzed 112 papers collected from different online sources, both from the academic field and from websites and blogs ranging from 2004 till mid-2022.\\
\textbf{Results}: a) There is no one single bullet-proof solution available for the protection of public spaces. b) From our analysis we found three major active fields for the protection of public spaces: Information Technologies, Architectural approaches, Organizational field. c) While the academic suggest best practices and methodologies for the protection of urban areas, the market did not provide any type of implementation of such suggested approaches, which shows a lack of fertilization between academia and industry. \\
\textbf{Conclusion}: The overall analysis has led us to state that there is no one single solution available, conversely, multiple methods and techniques can be put in place to guarantee safety and security in public spaces. The techniques range from architectural design to rethink the design of public spaces keeping security into account in continuity, to emerging technologies such as AI and predictive surveillance.
\end{abstract}



\begin{keywords}
Internet of Things \sep Cyber physical spaces \sep Public spaces \sep Cyber threat intelligence \sep Topic modelling \sep Topological data analysis \sep Protection public spaces \sep Smart city
\end{keywords}

\maketitle

\input{sections/1_introduction.tex}
\input{sections/2_background-relatedW.tex}
\input{sections/3_research_materials-methods.tex}
\input{sections/4_results.tex}

\input{sections/5_discussions.tex}
\input{sections/6_lesson_learned.tex}
\input{sections/7_limitations.tex}
\input{sections/8_conclusions-future_work.tex}
\input{sections/9_annex_I.tex}











\section*{Acknowledgment}
The work is supported by the EU H2020 framework programme, grant ``ANITA'' under Grant No.: 787061 and grant ``PRoTECT'' under Grant No.: 815356. Moreover, the authors thank the students Kaspar Raijmann, Vincent Fokker, and Thomas Meulenbroek from The Jheronimus Academy of Data Science (Netherlands) for the effort and support in analyzing, reading, and coding some of the documents and papers for this Systematic Literature Review. 

\bibliographystyle{cas-model2-names}
\bibliography{References}



\end{document}

%% file: sections/1_introduction.tex
\section{Motivation, Vision, and Goals}

Cyber-Physical Systems (CPS) are collaborative systems featuring devices and sensor networks interconnected with the surrounding physical world \cite{ColomboK14,Monostori2018}. CPS can be generally characterized as \emph{``physical and engineered systems whose operations are monitored, controlled, coordinated, and integrated by a computing and communicating core''} \cite{Rajkumar2010} and for these intensive interoperability characteristics---especially connected with physical-world counterparts---they pose several security challenges \cite{UllrichW15} but also play a key role in the interplay between the cyber- and physical places, where they both impact each other. For example, the cyber- element is providing applications of embedded computer and communication technologies for the automation of physical spaces (e.g., automotive, civil infrastructure, healthcare, protection of urban spaces) whilst information from the physical environment (road/patient/security conditions) are fed back into the cyber-counterpart \cite{TsigkanosKG16} to achieve a continuity of information able to yield data-driven policy-making and other key societal functions.

CPS typically embodies the following technological components:\

\begin{enumerate}
    \item embedded systems and Internet-of-things (IoT) technology, i.e., sensors, controllers, or actuators able to interact directly with the physical world;
    \item real-time systems or safety-critical systems that are able to perform actions in a predefined time;
    \item wired and wireless technologies able to communicate with each other sharing data and information (i.e., sensor networks); and,
    \item control-theoretical systems where most CPS observe and attempt to control variables in the physical world, such as sensors and actuators (i.e., pumps, valves, etc.), can be controlled remotely. 
\end{enumerate}

\begin{wrapfigure}{l}{0.5\textwidth}
    \centering
    \includegraphics[width=5.5cm]{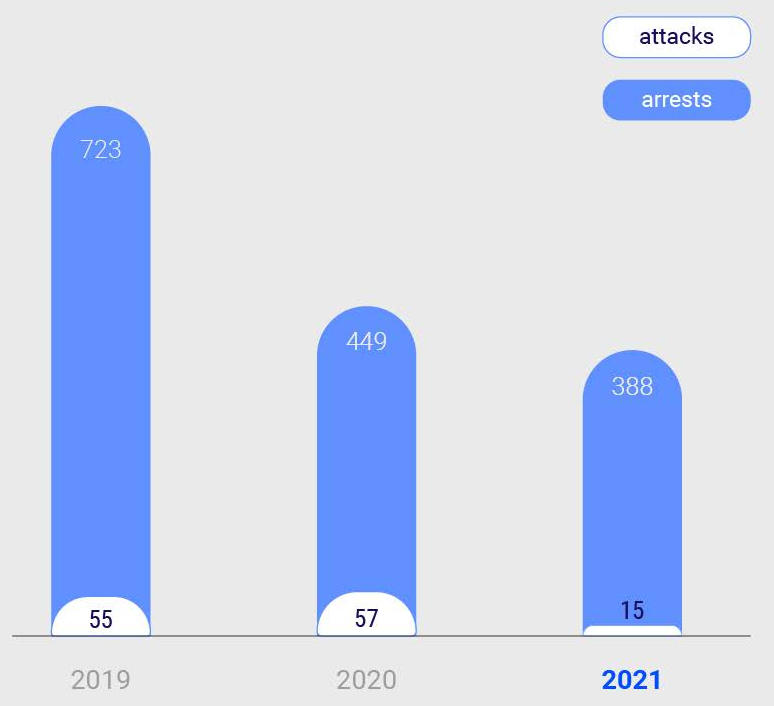}
    \caption{Number of failed, foiled or completed terroristic attacks in the time period between 2019 and 2021}
    \label{nattacks}
\end{wrapfigure}

By now, virtually all of the above key technologies have pervasively and deeply penetrated the tissue of our modern societies \cite{SimonJE14}, that is, the \emph{cyber-physical space} element in the CPS equation, everything around and interplaying with the CPS itself \cite{TsigkanosKG16}. Indeed CPS technologies constitute an ideal candidate for the protection of high-risk targets in urban areas like squares, streets, avenues, and concerts in the scope of public events. Such protection is nowadays even more dire: terrorist attacks pose one of the major concerns for any nation \cite{Clutterbuck1990}, and, in some areas of the world, civil servants and municipalities have to confront such threats almost daily. Even focusing on the EU, it can be witnessed that in 2015 alone, terrorism hit the headlines six times, specifically, the attacks in Paris, including the headquarters of the Charlie Hebdo and the Bataclan Theater, resulting in 138 people killed and 413 injured \cite{boutry_2019,rodionova_2016}. Recently in 2021, a female police employee has been killed in a knife attack at a police station in Rambouillet~\cite{Rambouillet_attack}. In Murcia in 2021 the town of Torre Pacheco faced a multiple hit-and-run attack on the terrace of a bar, resulting in two deaths, one of them the driver of the car, and several injured~\cite{murcia_attack}. In 2022, Germany, 28-year-old pulled a knife on four passengers attempting to murder them~\cite{germany_attack}. Vaughn Dolphin from Walsall Wood Road in Aldridge, Walsall, was arrested on 27 June 2022 at an address in Cheshire on suspicion of terrorism offenses and in possession of explosive material and manuals on how to make improvised explosives and firearms~\cite{walsall_man}. Analysts deemed the ``intelligence failure'' factor as one of the most neuralgic criticalities of such attacks \cite{rodionova_2016}, but unfortunately, these post-terror analyses and key findings did not (yet) yield serious improvement in intelligence \cite{bbc_news_2016,bbc_news_2017,bbc_news_2019,hache_2016,kommenda_holder_clarke_levett_cage_ulmanu_sheehy_guest_2017,report_evolution23}.
Europol reports that only in 2021 388 individuals have been arrested in Member States for terrorism-related offenses. Europol highlighted how terrorist attacks in the EU sharply decreased in 2021. This can be attributed to a change in classification of incidents by some Member States from terrorist to extremist attacks \cite{europol_2022}. Figure~\ref{nattacks} shows the number of attacks in Europe from 2019 to 2021 \cite{europol_2022}. 

Germany recorded the second highest number of terror attacks for the region in 2021, and has the fourth-highest overall impact of terrorism in Europe.  Turkey remains the country most affected in Europe~\cite{global_terrorism22}.
In this regard, CPS plays a key role in protecting public spaces and the overall security enhancement of the next generation of smart cities. CPS are IT systems introduced in the applications of the physical world and can monitor and control the physical world at runtime. Hence, CPS can help authorities monitor, prevent, and mitigate possible terrorist attacks, pickpocketing, harassment, public violence, or mob control.





Thus far, the prevention, protection, and preparation against terrorist attacks have been largely studied in a rather siloed, mono-disciplinary manner, e.g., from the domain of criminology, law, and information technology. Much less emphasis has been placed on the cyber-physical dimensions of such attacks that intrinsically ingrain a multi-disciplinary dimension. This survey addresses this shortcoming by providing a systematic multivocal literature survey comprised of an updated, comprehensive and timely overview of state of the art in counter-terrorism cyber-physical systems, hence aimed at the protection of \emph{cyber-physical spaces}. 

Our investigation lens is \emph{multivocal} since it accounts for both grey (e.g., blog posts and white papers) and white literature (e.g., journal and conference papers) on the matter with a systematic multivocal literature review (SLR) \cite{Garousi:2013:EUQ:2460999.2461003,KITCHENHAM20097}. The analysis builds further on the---in several cases outdated---observations and outline of information technologies already identified previously by Popp et al. in 2004 \cite{popp2004countering} and offers a more in-depth overview for computing practitioners interested in deepening their knowledge over this neuralgic area of research for societal impact. Furthermore, our overview offers a much needed birds-eye view over the state-of-the-art in the targeted domains for more instrumented and precise decision- and policy-making.

Understanding the interactions of the cyber-physical system with the physical world would bring about many practical and theoretical benefits. For example, practitioners and security officers can design newer attack-detection algorithms. Similarly, policy-makers could better understand the consequences of a physical attack and can design a smarter city~\cite{GujratiZS18,policing_smartcity,Zhang20}, able to better prevent and mitigate attacks. We envision, with such cyber-physical systems and spaces, a city that would indeed become a smarter urbanized object, proactively providing alerts and being protective against any threat even in a participatory fashion, e.g., including citizen-science counterparts \cite{olson2012cloud,app12041863} for specific purposes (e.g., proactive threat detection \cite{ParkYH09}). Moreover, citizens might improve their own awareness and capabilities when dealing with such threats and connected risks as they manifest. Our study offers a first overview to focus on technologies and solutions from the literature that propose systems to enhance the protection of cities' external and public environments in search of the aforementioned vision.


The main research question this research aims to answer is \emph{``What are the best practices and technologies nowadays available to protect urban cyber-physical spaces?"}. Hence, the main contributions of this paper are: (a) an overview of state of the art in the current CPS based on a systematic literature review (SLR); (b) the preparation of quantitative indicators for further risk assessment and research in underdeveloped CPS areas revealed; (c) a comprehensive overview of techniques, prevention measures, and, methods of assessment for targeted information technologies; (d) best practices adopted by governments and private organizations; and lastly, (e) a comprehensive replication package that can be used either to proceed the proposed research as well as for practitioners to get to grips with the proposed technologies and methodologies in our synthesis\footnote{Replication package available at: shorturl.at/hoyJT}.


The remainder of this paper is organized as follows. Section~\ref{b&rw} provides an overview of related works in the field of protection of public spaces. Further on, Section~\ref{research_matherials_methods} elaborates on the research design behind this study. Subsequently, in Section~\ref{results} the results of our analysis are presented, after which Section~\ref{discussions} and \ref{limitations} provide discussions and limitations of the research, respectively. Finally, Section~\ref{conclusions} discusses results, draws conclusions, and sketches a roadmap for future work. 


%% file: sections/2_background-relatedW.tex
\section{Related Work}
\label{b&rw}
Nowadays, cities are not well equipped to provide a high level of security to the citizens against terrorist attacks. Europol in \cite{europol_2022} provided some trends regarding the last years. Newspapers are full of news regarding terrorist attacks, and cities like London \cite{londonattack21}, Liverpool \cite{liverpoolAttack}, The Hague \cite{theHagueAttack} have faced terrorism in the latest years. It is hence, of vital importance to provide and introduce cyber-physical systems to enhance the overall capabilities of cities, making them smarter, more resilient, and able to face any type of event that can mine the safety and security of citizens.

The systematic literature review presented in this work is a partial replication inspired by the work in Popp et al. \cite{popp2004countering}. 
On the one hand, our results offer a much more elaborated taxonomy and connected results about the best practices and information technologies related to CPS to protect urban spaces. 
On the other hand, Popp et al. \cite{popp2004countering} defined 15 different information technologies that apply to the protection of urban space. To verify the existence of these technologies in literature and to exploit possible newly emerged technologies, the systematic literature review reported in this paper offers analyses of the state of the art over current CPS and information technologies for the protection of urban spaces. Furthermore, beyond Popp et al., the review---as a major novel contribution---also takes into consideration sources from the open web (e.g., blogs, news articles, and product publications), checking for their quality, given the novelty of specific technologies and since many such technologies emerge in the private and corporate domains with little to no penetration in the state of the art.

As previously mentioned, the first work to address the objectives behind this research is from Popp et al.~\cite{popp2004countering}. The authors provide an analysis of what was currently known about countering terrorism through information technologies. The main goal of this paper is to offer an updated, more in-depth analysis stemming from the same concepts and definitions \cite{popp2004countering}. Popp et al. identify 15 information techniques most vital to preventing and countering terrorism attacks. The research focused on collaboration, analysis, and decision support tools, foreign language speech and text analysis tools, and pattern analysis tools to analyze and solve complicated terrorism-related problems more efficiently and effectively. The study concluded that the three core IT areas were not merely close to the knowledge available on countering terrorism, and many other information technologies are crucial for successfully countering terrorism \cite{popp2004countering}. To validate the advantages of these techniques, this study aims to map the available knowledge, going beyond the cataloging of such knowledge but delving into a deeper layer of analysis on top of what was known.


An example of CPS' study for the protection of public spaces is in the work of Zab\l{}ocki et al. \cite{zablocki2014intelligent}. In 2014, Zab\l{}ocki et al. surveyed the latest state-of-the-art intelligent video surveillance systems and their most desirable features and characteristics. After analyzing and evaluating the video surveillance systems in several categories, like object detection and detecting and identifying abnormal and alarming situations, the study concluded that there are several challenges still to be overcome. These challenges involve legal, and privacy protection, difficulties with object movement analysis for occlusion handling, and synchronization of multiple camera views during real-time system operation \cite{zablocki2014intelligent}. Similarly to the previous one, we have an analysis of the smart surveillance monitoring system in \cite{Tavagad2016SurveyPO}. 

However, the proposed results simplistically inspire the use of a low-cost single onboard computer Raspberry Pi to reduce the cost for a smart camera system. As well, looking at being cost-effective, the solution proposed in \cite{Raju2015SmartVS}. The idea is to use Android phones in conjunction with Open CV technology to create a system where the CPS admin can have 24/h per day under the control of the system directly through his phone. If from above, we saw hardware-related CPS papers in \cite{Sreenu2019}, and in \cite{8926351} we have the usage of video surveillance systems for deploying crowd monitoring systems. The former proposes a review of papers on intelligent surveillance video analysis techniques showing a wide variety of applications. In the paper, crowd management's application areas range from object and group tracking, trajectory filtering, abnormal behavior detection, human subject detection and discrimination, and dangerous activities. 

Instead, the latter proposes automated video analytics for crowd monitoring and tracking to enable significant manpower savings in CCTV monitoring. The work relays on the usage of \textit{Python} and \textit{OpenCV library} comparing a Convolutional Neural Network (CNN) approach to a Background Subtraction (BGS). CNN showed to be the best method suitable for deployment for crowd monitoring in public spaces. 

Unlike the study mentioned above, we have the literature review from Sadiku et al. in \cite{Sadiku2017CyberPhysicalSA}. In the paper, they provide a brief introduction to CPS and its applications. We have CPS in the manufacturing, healthcare, transportation, agriculture, and computer networks concluding.

Part of the ecosystem above is also sensors technologies. A Cyber-Physical Sensor System (CPSS)~\cite{cpss} consists of a computing platform equipped with wireless access points, sensors, and actuators~\cite{cpss}. In a cyber-physical system, CPSS typically plays a fundamental role since they are responsible for collecting data from the physical word and are able to perform real-time activities based on the process algorithm \cite{Berger2016CharacterizationOC,8541316,10.1145/3399715.3399826,https://doi.org/10.1002/smr.2511}.

Finally, the domain of CPS technologies includes Internet of Things (IoT), Industrial Internet, Smart Cities, Smart Grid, and ``Smart''-* (e.g., Cars, Buildings, Homes, Manufacturing, Hospitals, Appliances), for the sake of space we mentioned just a small group of CPS examples for the protection of public spaces. 

From the overview of the related work, we find that most of the time, other publications focus on a specific type of technology (hardware or software) to provide specific information on the current state-of-the-art. While, we aim to provide a broader comprehensive overview of available approaches for the protection of urban areas. This literature review aims to build a document for academics, practitioners, LEAs, and municipalities where to find hints on how to protect urban areas and the current lack in the field. For the insiders, our findings aim at providing an overview of the state of the art and finding possible collaboration opportunities. While for outsiders, our findings represent an easy document to read to find approaches and technologies for the protection of public spaces, but also hints and ideas of possible interactions with other shareholders. The overall document aims at building a milestone in building interaction among LEAs, municipalities, practitioners, and academics. To the best of our knowledge, this literature review represents the first attempt to bring together multiple subjects around the shared problem of CPS for the protection of urban spaces providing a guideline among the different types of approaches.

\begin{framed}
\textbf{Research Objective}. The present SLR aims to focus on CPS technologies for the protection of cyber-physical enabled and supported public spaces. To the best of our knowledge, this is the first multi-vocal SLR spawn from the results and conclusions of Popp et al. in \cite{popp2004countering} that makes a coherent overview of the aforementioned technologies and best practices for the protection of urban areas. 
\end{framed}


%% file: sections/3_research_materials-methods.tex
\section{Research Materials and Methods}
\label{research_matherials_methods}
This section provides an overview of the methods used for obtaining the white and grey literature and the research questions and means employed to obtain the final results.

\subsection{Research Questions and Approach Overview}
The methodology used to attain our results is described in the following paragraphs and is tailored from our previous work \cite{SOLDANI2018215}. More specifically, we seek to address the following key research question:\par
\begin{center}
    \textbf{MRQ.} \emph{What are the best practices and CPS technologies available nowadays for the protection of urban spaces?}
\end{center}

The aforementioned question can be rephrased into the questions reported here below:

\begin{enumerate}[start=1,label={SRQ\arabic*.},wide = 0pt, leftmargin = 3em]
    \item To what extent do new technologies effectively enhance the protection of public spaces?
    \item What are the newest approaches available for the protection of urban areas?
    \item What is the extent of agreement among white and grey literature, and what are the research gaps and overlaps?

\end{enumerate}

The research questions were elicited from 7 focus-group workshops held at five municipalities across the EU and featuring local law-enforcement agencies (LEAs); the focus of the workshops was kept closed and about vulnerabilities regarding each city involved. Specifically, this phase of the study featured the following organizations and citizenships: (1) Inspectoratul General al Politiei Romane (RO); (2) Lietuvos Kibernetiniu Nusikaltimu Kompetenciju Ir Tyrimu Centras (LT); (3) Gemeente Eindhoven (NL); (4) Ayuntamiento De Malaga (SP); (5) Forum Europeen Pour La Securite Urbaine (FR); (6) Dimos Lariseon (GR); (7) Vilniaus Miesto Savivaldybes Administracija (LT); (8) Municipiul Brasov (RO); (9) Ministerio Del Interior (SP). 

The general aim of said workshops---each held in the involved municipalities of Brasov (RO), Vilniaus (LT), Eindhoven (NL), Malaga (SP), Lariseon (GR)---was to identify the weaknesses in the current security measures in a specific area chosen by the municipality and in reference to a possible terrorist attack, as well as to increase the situational consciousness of the municipal staff and the relevant stakeholders responsible for the public security. Specifically, during the workshop we had the possibility to collect information regarding three research areas: \emph{a} the type of technologies in place in a pre-identified public space; \emph{b} lacks and problems related to the security of a public event in the pre-identified public space; \emph{c} the solutions proposed by municipalities and LEAs to solve specific issues connected to the security of a pre-identified public area.

The workshop was part of the PRoTECT project\footnote{PRoTECT: \url{https://protect-cities.eu/}}, the specific results from the workshop, the technologies currently in place, the proposed approach and solutions, and the public areas involved in the project can not be disclosed for security reasons. However, some details are available on the page available in the footnote\footnote{\url{https://protect-cities.eu/protect-project/public-deliverables/}} where are shared all the public deliverables.

For each workshop, 3-5 stakeholders with 3+ years of expertise in the relevant fields (law enforcement, combating terrorism, maintaining and re-establishing public order, operational centers, pyrotechnic control, explosives and management of emergencies) were gathered to discuss and bring their professional expertise on different security aspects regarding the analyzed Public Space of Interest (PSOI)~\cite{PSOI, Carmona2019}. 

The research questions from our SLR resulted from the most prioritized items based on the daily work carried out by LEAs and municipalities in order to guarantee safety and security in public spaces. 

Our SLR takes those RQs as a basis for investigation and covers all the aspects and technologies of protection of public spaces in said RQs, namely, (1) technologies available to protect urban spaces, (2) designs, methods, and approaches to mitigate, avoid or prevent attacks, (3) techniques for crime security, analysis, and management.

More in detail, in terms of SRQ1, we aim to understand the available technologies and the upcoming technologies for the protection of public spaces. Conversely, SRQ2 focuses on approaches, methodologies, and new techniques available for the protection of urban spaces. We thought of multiple governance methodologies, aiming to offer a diversified outlook as much encompassing state of the art and the potentials thereof. Lastly, SRQ3 identifies and analyses the gaps between the white and grey literature to understand how feasible the approaches suggested by academics are.  

In order to address the research questions above, this work adopted the well-known Systematic Literature Review (SLR) research approach \cite{kitcha2007}. More specifically, the approach is detailed as follows.

First online literature is sampled to provide an overview of the techniques, technologies, and tools used for the protection of urban spaces. 
Subsequently, unsupervised several analyses were incepted, including Machine-Learning analysis techniques used explicitly for topic modeling and literary analysis \cite{OnanKB16,WilliamsB18} as well as content analysis \cite{krippendorff80,HsiSha05} used to gather indicators for (1) further assessment of the technologies and information systems found in literature as well as (2) identification of any existing gaps in the scientific literature.



The remainder of this section outlines all the above phases and the techniques adopted within them in more detail.

\subsection{Sample Selection \& Control Factors}
\label{sample_selection}

All of the papers used for this research either belonged to white literature or grey literature found on the Internet. White literature, defined as research published by an established scientific organization, was enriched with articles and papers scraped from governmental websites to find further research. Most grey papers were found on websites dedicated to sharing information concerning information techniques and private companies outlining their best practices. Selection criteria are applied to the white papers to limit the span of the research. The inclusion criteria consist of:

\begin{table}
\centering
\begin{tabular}{lp{12cm}}
\hline
	{\bf Case}
	&
    {\bf Criteria}
    \\
\hline
\rowcolor[HTML]{EFEFEF} 
	Inclusion 
    & 
	$\mathsf{i}_1$) The study discusses methods or technologies to address the topic.\newline
	$\mathsf{i}_2$) The study discusses the challenges around the topic close to our RQs.\newline
	$\mathsf{i}_3$) The study addresses know-how, guidelines or best practices on the topics in our RQ by directly-experienced LEAs, municipalities or practitioners.
	\newline
	$\mathsf{i}_4$) The study reports a case-study of urban space attack incidents or approaches.
    \\
 	Exclusion 
    &
    $\mathsf{e}_1$) The paper does not discuss sufficient details on implementation of practices, methods or tools for protection of urban spaces.
    \newline
	$\mathsf{e}_2$) The discussed topics are not explained/evaluated by the paper.\newline 
    $\mathsf{e}_3$) The research paper does not discuss scope and limitations of the proposed frameworks, methods, tools, solutions, guidelines.\newline
 \\
\hline
\end{tabular}
\caption{Inclusion and exclusion criteria for sample selection.}
\label{tab:criteria}
\end{table}

The inclusion criteria ($\mathsf{i}_1 - \mathsf{i}_4$) represent the parameters are used in order to identify the targets of this study. On the other side, the exclusion criteria are used to exclude papers that do not match the defined targets, i.e., $\mathsf{e}_1$ to exclude studies with poor design/implementation details or $\mathsf{e}_4$ in order to exclude studies that do not examine limitation and impact of the proposed solution. 
Every paper used in this research needs to satisfy all the inclusion criteria to be added, while it will be excluded if it satisfies at least one exclusion criterion. 
Based on the criteria stated in Table~\ref{tab:criteria}, 83 papers belonging to white literature and 29 grey articles were selected. For the screening process, not only the inclusion and exclusion criteria were used but also some quality control factors:
\begin{itemize}
    \item Articles must be written and published after 2004;
    \item Articles must be written in English;
    \item Articles must be written or published by a trustworthy source;
\end{itemize}
The research of Popp \cite{popp2004countering}---published in 2004---provides the foundation of this research. In order to guarantee the most up-to-date information, only articles after 2004 are included in this research. The next criterion is that all articles used must be written in or translated into English. The choice of using only the English text is to ensure the quality of the content and allow it to be replicable. Finally, since not all information concerning information technologies in the context of countering terrorism or the public's safety on the internet can be considered to be accurate, only papers and articles published by trustworthy sources will be used in this research. A trustworthy source is a computing literature library like ACM Digital Library; IEEEXplore; Wiley Interscience; Elsevier Scopus; and Bibsonomy for the academic literature. Meanwhile, for the grey literature, we double-check our source to find the same news on multiple channels like digital newspapers and technological magazines, and blogs to have a higher certainty of the trustworthiness of the considered reference. Hence, we used less strict criteria than the white papers, and Google search was the main engine for this type of online research.  

\subsection{Search strategy}

To find articles, we started using the keywords from Table~\ref{table_keywords}, which were coded directly from the interview and workshop transcripts elicited through our preliminary pilot study (see sec. \ref{research_matherials_methods}, Par. 3 and following). The sources were screened against the practitioners that originally participated in a preliminary pilot study; this way, we were able to guarantee that the sources from which the primary studies were drawn were reliable sources of truth. Each query produced a considerable amount of articles and journal papers. We handle the process of going through titles and abstracts to include or exclude papers based on the criteria explained in Table~\ref{tab:criteria}.
To acquire our articles, we proceeded through three search rounds. The first round produced 96 articles, and a second round produced 30 papers. In total, we collected 126 articles covering the years range 2004-2020. We then proceeded with a third round that produced 8 more papers and covering the years between 2020-2022 to have the most updated SLR for a total of 134 papers.
After the articles were obtained, the selection was narrowed to industrial, government, and non-governmental paper researches (e.g.,~blog articles, white papers, magazines) published from 2004 until mid 2022. Google (primary) and Bing are the two search engines mainly used to scrape \textcolor{blue}{the grey literature}. \textcolor{blue}{We opted for Google and Bing due to their capabilities of indexing pages. We used the keywords from Tab.~\ref{table_keywords} to find blog articles, magazines, and industrial papers.} At the same time, to cover for white literature appropriately, queries were running in typical and most common computing literature libraries, namely: (1) ACM Digital Library; (2) IEEEXplore; (3) Wiley Interscience; (3) Elsevier Scopus; (4) Bibsonomy (5) Google Scholar; and (6) and Science Direct. These digital libraries were drawn from previous and related surveys i.e., \cite{CASCAVILLA2021102258,10.1145/3369052}.
References from articles found and the corresponding journals were also used to gather information in addition to the stated online library databases. 134 papers have been analyzed \textcolor{blue}{by all the authors plus three students (mentioned in the acknowledgment, and working as students project assistant)}, and applying once again the inclusion and exclusion criteria we came up with a final number of 83 academic and 29 non-academic articles included in this SLR.
In order to address the \textit{Inclusion} and \textit{Exclusion} criteria mentioned above, the relevant keywords to be used to find the white and grey literature were defined. A keyword search is a search type that checks for matching documents involving one or more words specified by the user. These keywords need to discriminate against the most relevant research among the scientific databases \cite{sari2019systematic}. Keywords, displayed in Table~\ref{table_keywords}, are those used for searching in the databases listed above. Multiple keywords were combined to enhance the search results by filtering out irrelevant articles. 
A list of combined queries is provided here below:
\small{
\begin{itemize}
    \item $(\texttt{information*} \wedge \texttt{management*} \wedge \texttt{filtering*})$
    \item $(\texttt{knowledge*} \wedge \texttt{management*}) \vee ( \texttt{context*} \wedge \texttt{management*})$
    \item $(\texttt{predictive*} \wedge \texttt{modelling*})$
    \item $(\texttt{categorization*} \vee \texttt{clustering*})$
    \item $(\texttt{event*} \wedge \texttt{detection*}) \vee ( \texttt{event*} \wedge \texttt{notification*})$
    \item $(\texttt{geospatial*} \wedge \texttt{detection*} \wedge \texttt{exploitation*})$
    \item $(\texttt{workflow *} \wedge \texttt{management*})$
    \item $(\texttt{semantic *} \wedge \texttt{Consistency*}) \vee ( \texttt{resolving *} \wedge \texttt{terms*})$
\end{itemize}
}

In Table~\ref{table_keywords}, the list of the keyword used to build our queries for this literature review. 

\begin{figure}[h!]
    \centering
    \includegraphics[width=12cm]{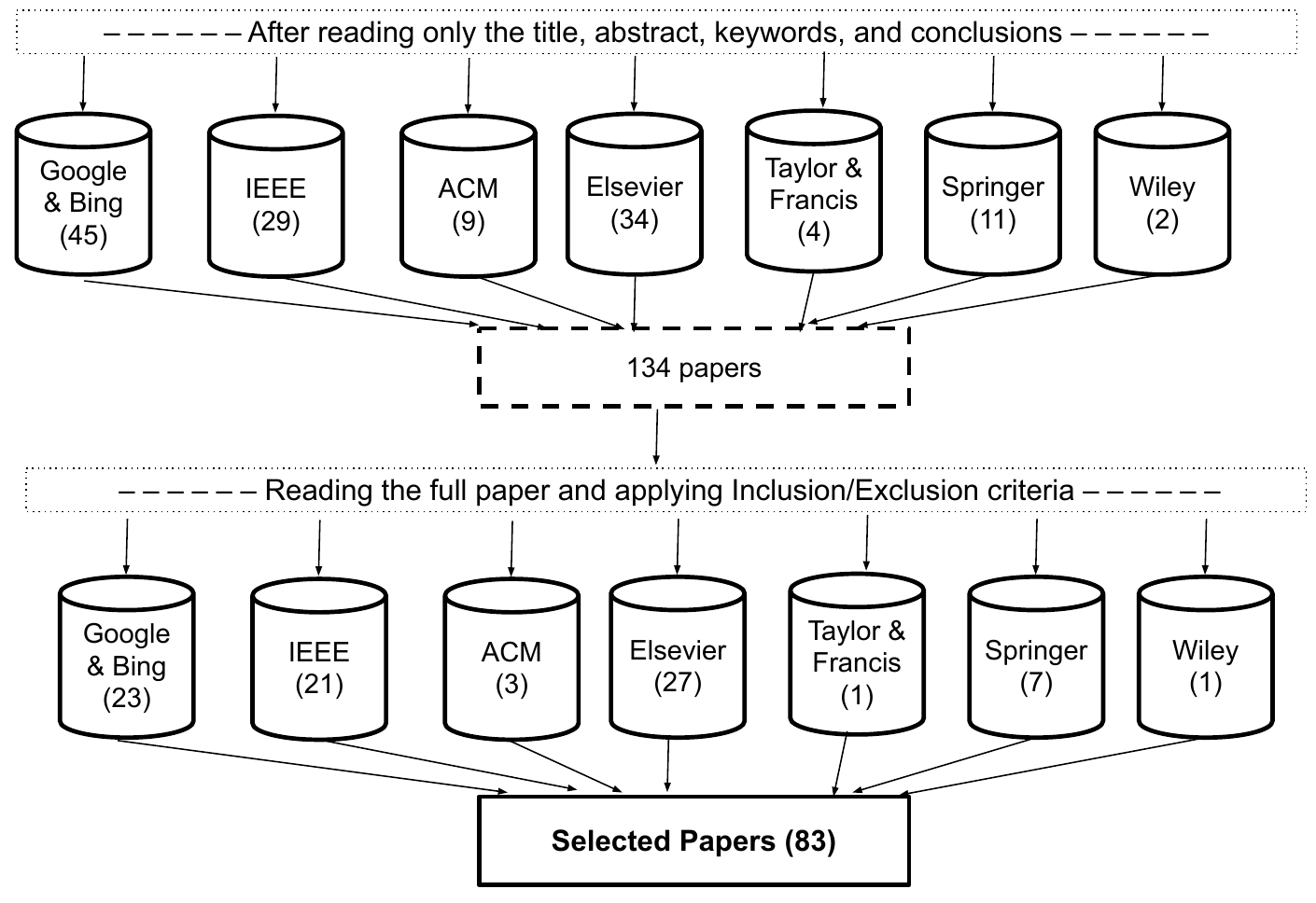}
    \caption{Systematic literature review selection process.}
    \label{fig:search_strategy}
\end{figure}

\textcolor{blue}{Below, Fig.~\ref{fig:search_strategy} presents an overview of the review process, showcasing the number of identified papers and the corresponding selected papers. To ensure comprehensive coverage, we employed both the Google and Bing search engines to include papers that may not have been published yet, such as those from arXiv, as well as those from lesser-known libraries.}

The target in question is not necessarily software engineering, but more generically information and software technology related, previously used, or evaluated in the context of cyber-threat intelligence—intended as the discipline and practice of diagnosing cyber-threats, by any means, i.e., analytical or otherwise—along with the typical context in which such efforts are to be found, that is, law-enforcement, systems surveillance and their general security and safety governance. The reason for selecting such terms and queries is drawn directly from the stakeholder interviews and confirmatory workshops we held with the (Cyber-)Law-Enforcement Agencies (LEAs)—primary targets for the exploitation of our results—that were involved in our study and which represent a unique ground truth beyond question.

\begin{table}
\centering
\begin{tabular}{lll}
\hline
\textbf{Keywords:}   &                         &                       \\ \hline
Biometrics;          & Filtering;              & Publishing;           \\
Categorization;      & Geospatial;             & Resolving terms;      \\
City;                & Government;             & Safety;               \\
Clustering;          & Information management; & Searching;            \\
Context management;  & Information technology; & Secure;               \\
Counter terrorism;   & Infrastructure;         & Semantic consistency; \\
Crime;               & Knowledge management;   & Terrorism;            \\
Data;                & Machine-Learning;       & Urban;                \\
Database processing; & Predictive modelling;   & Video processing;     \\
Event detection;     & Prevention;             & Visualization;        \\
Event notification;  & Public space;           & Workflow management.  \\ \hline
\end{tabular}%
\caption{Table of keywords used.}
\label{table_keywords} 
\end{table}

\subsection {Data extraction}
For the extraction of the data, a mixed-method analysis approach was used to obtain qualitative and quantitative data at once \cite{johnson2004mixed}. For this, a thematic coding content analysis method was applied to generate themes from the involved data. The thematic coding was done in a six-phase analysis method, including:
\begin{itemize}
\item \textbf{Familiarization}. We started reading the found papers in order to become familiar with the found data.
\item \textbf{Generating the initial codes}. After we became familiar with the data, we started the coding stage. In this phase, it is important to isolate those phrases, sentences, and paragraphs related to our topic. We analyzed all the papers and extracted phrases, sentences, and paragraphs to create clusters of themes.

\item \textbf{Create the initial themes}. In this phase, we revised the codes clustering them together if there were similar meanings or if we found relationships among them. This stage helped us identify patterns among codes.
\item \textbf{Review the initial themes}. In this phase, we took the themes and began to review them against the data.
\item \textbf{Name and define the themes}. In this phase, it is important to use proper labels to create meaningful and comprehensive names for the different themes involved. We revised our themes to give them more comprehensive names.   
\item \textbf{Write the final report}. After the definition of the themes naming them, we started writing our final report as defined in \cite{braun2018thematic}. 
\end{itemize}


Finally, to address SRQ1 and SRQ2, we operated a machine-assisted topic modeling and analysis exercise supported with a thematic coding. We chose the Latent Dirichlet Allocation (LDA) approach since it is generally the most popular and effective topic modeling technique. With LDA from our dataset of papers, we extract human-interpretable topics (topics or themes are a group of statistically significant words within a corpus, we will use the terms interchangeably~\cite{LDA_gensim_sklearn}) characterized by those words that are most strongly associated with. More specifically, we used Latent Dirichlet Allocation (LDA) to detect emerging themes in our textual data.

\subsubsection{Thematic Coding}

Thematic coding was adopted to get a baseline understanding of the state of the art information technologies to protect urban space.
The list of themes is partially based on the work made by Popp et al. in~\cite{popp2004countering} and dated 2004. In the work by Popp et al. are identified some of core IT areas considered crucial for counterterrorism. With our proposed literature review, we intend to improve this list of technologies to resemble the current state of the literature on information techniques. One researcher and three student assistants have been involved in the process of selection, analysis, labeling, and finding of emerging themes. The first round concerned the usage of the keywords from table~\ref{table_keywords} to find related publications, articles, and grey literature from the specified academic and non-academic channels. The first round produced 96 academic papers for the white literature, and with a second round of additional 30 research papers, for a total amount of 126 academic articles. After the analyses process, the papers were reduced to the final number of 83, while for the grey literature, we processed 29 articles. During the reading and analysis process papers have been removed since were not providing specific type of technologies or approaches for the protection of public spaces.
The selected set of articles obtained by the article search was each individually analyzed, labeled, and the most important parts were subtracted and stored in an overview with the emerging theme that arose from the analyzed text. This overview served as a total collection of essential parts of the analyzed text with their concurrent themes and articles. 
To attain the findings, we based our Thematic coding approach on the work in \cite{Qualitative_Research}, where all the steps are presented and explained. The analysis consist on six steps, mainly: \emph{(1) Familiarization}, \emph{(2) Generating the initial codes}, \emph{(3) Create the initial themes}, \emph{(4) Review the initial themes}, \emph{(5) Name and define the themes} and lastly \emph{(6) Write the final report}.
The thematic coding aimed to code as consistent as possible all the papers in the white and grey literature. The results of the manual coding exercise are listed in Table~\ref{tab:table_themes}.

\begin{table}
\centering
\begin{tabular}{ll}
\hline
\textbf{Code} & \textbf{Definition} \\ \hline
METH          & Methods for the risk assessment new data sources    \\
PRED          & Prediction model for prevention measures          \\
ASSES         & Modeling for assessment. How is the given model evaluating a state         \\
FEAS          & Feasibility of the posed technique              \\
BPGO          & Best practices  applied by governmental organizations        \\
BPOP          & Best practices  applied by other parties than governmental organizations         \\
TECH          & Prevention technologies/tools/designs                                      \\
ARIN          & Artificial intelligence (machine-learning)  
          \\ \hline
\end{tabular}%
\caption{Table of the emerged themes during manual coding.}
\label{tab:table_themes} 
\end{table}

\subsubsection{Topic Modelling and Topological Analysis}
We chose the Latent Dirichlet Allocation (LDA) \cite{Bas03} approach for our topic modeling analysis since it is the most popular and generally the most compelling topic modeling technique. With LDA, we can extract human-interpretable topics from our dataset of papers. Each topic is characterized by those words that are most strongly associated with it. Hence, we selected log-likelihood as our measure of clustering quality, following typical approaches from state of the art (\cite{AgrawalFM16}).

Besides, we used the genetic algorithm Differential Evolution to tune LDA hyperparameters alpha and beta, as suggested by Agrawal et al. (\cite{AgrawalFM16}). For the sake of space, we can explain the hyperparameters by saying that \texttt{alpha} is a parameter that controls the prior distribution over topic weights in each document, while \texttt{beta} is a parameter for the prior distribution over word weights in each topic. For the sake of space, we refer the reader to the explanation provided in \cite{topicmodellingguide}.
Moreover, to conduct all the pre-processing phases and analyses, we exploited the NetCulator bibliometric analytics tool\footnote{\url{https://www.netculator.com/}}, which supports LDA and several similar natural-language analyses and clustering techniques and tools of our design, featuring Python and the python LDA package. 
To further analyze our dataset, we apply structural, topological Analysis \cite{KwittHNLB15}. Topological data analysis (TDA) refers to statistical methods that find structure in data. We decided to apply both TDA and LDA to study and visualize the overlap between these two analyses. Indeed, both analysis work with clusters of words. However, from one side, LDA produces a collection of documents that the algorithm has grouped based on a certain topic. Differently, TDA finds a connection among these clusters and visualizes the shape of the network behind these topics and how they are connected. Hence, we expect to have overlap among LDA and TDA analysis, where the former show topics, and the latter shows the interconnection among them \cite{LiK10}. 

%% file: sections/4_results.tex
\section{Analysis of the Results}
\label{results}
This section elaborates on the results of the analysis of this paper. First, descriptive statistics of the examined papers are reported. An overview is given concerning the year of publication, type of study, and the venue in which the article was published. Subsequently, we provide and discuss the results of the LDA topic modeling. While to study the topological shape of our dataset, we decided to apply Topological Data Analysis (TDA). To conclude, we performed t-Distributed Stochastic Neighbor Embedding (t-SNE) analysis, a statistical method for visualizing high-dimensional data in a two-dimensional map.

\subsection{Descriptive Statistics}
\begin{figure}[b]
    \centering
    \includegraphics[width=8cm]{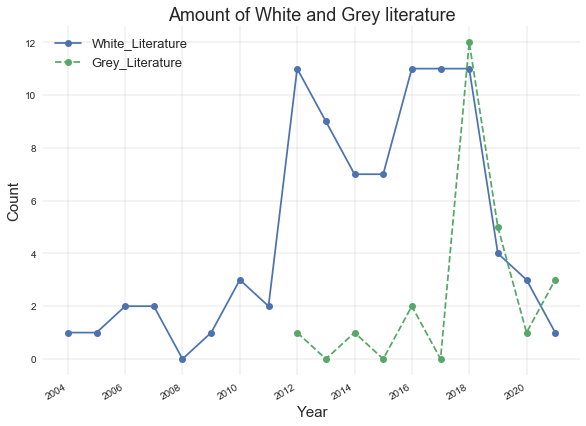}
    \caption{Amount white literature per year of publication.}
    \label{fig:white_literature} 
\end{figure}

The white literature included in this research range between the year 2004 and mid 2022. In order to limit the scope of this research, a minimum boundary was set. As mentioned in Section~\ref{sample_selection}, the literature included in this research should be published after the year 2003. An emerging trend line is displayed in Fig.~\ref{fig:white_literature}. On average, more recent literature has been found and included in this research. In Fig.~\ref{fig:white_literature} the papers from 2022 are added up to those of 2021 since, at the end of collecting the papers, 2022 was only advanced halfway. It is to be noted that our analysis approach is based on previously and successfully published studies with similar depth and scope, primarily~\cite{CASCAVILLA2021102258,10.1145/3369052}.

\begin{figure}\centering
\subfloat[Type of studies within white literature]{\label{fig:type_white_literature}\includegraphics[width=.40\linewidth]{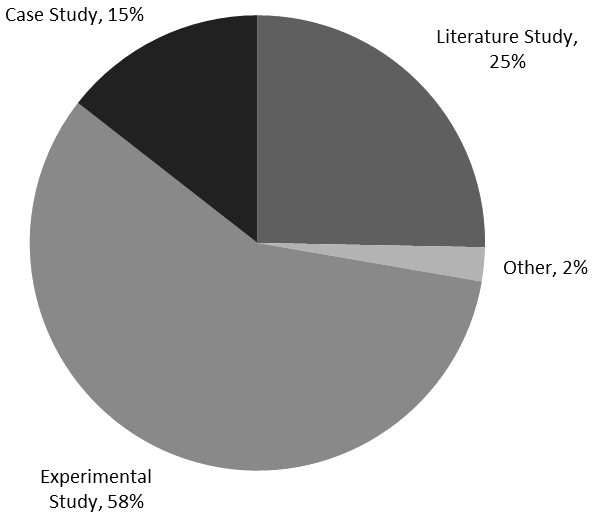}}\hspace{-1mm}
\subfloat[Venues used for publication of white literature]{\label{fig:venues_white_literature}\includegraphics[width=.55\linewidth]{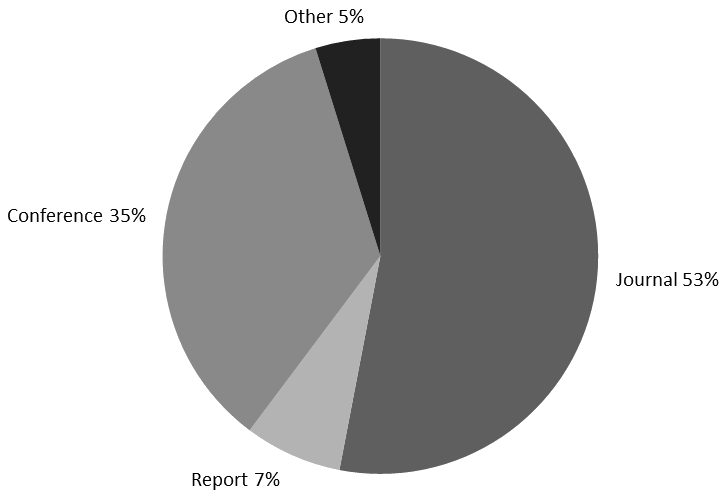}}\par 
\subfloat[Type of websites used for grey literature]{\label{fig:type_of_web_sorces}\includegraphics[width=.45\linewidth]{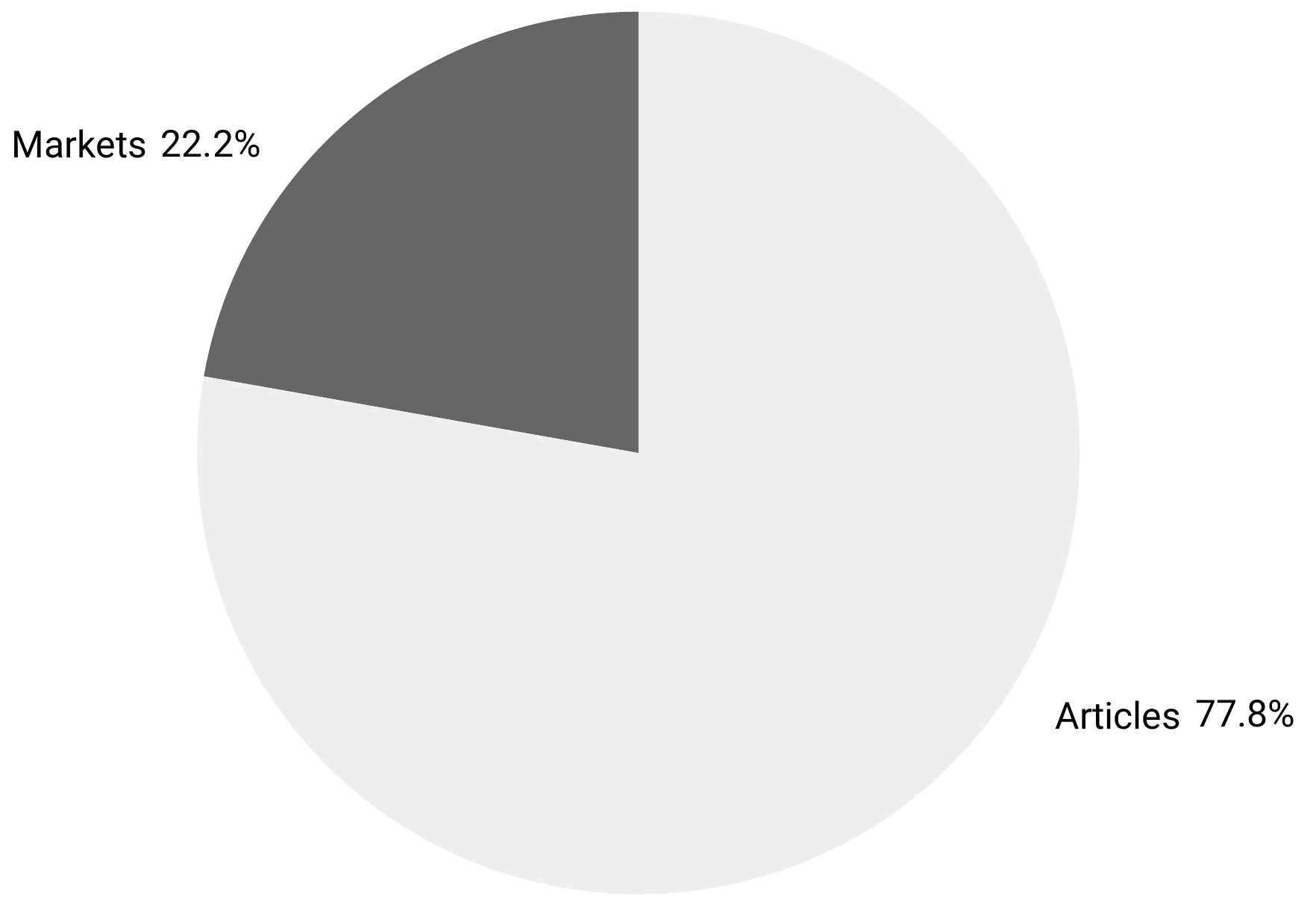}}
\caption{a white (top) and grey (bottom) literature dataset: descriptive stats for types of studies, venues, and online sources.}
\label{fig}
\end{figure}

In Fig.~\ref{fig:type_white_literature}, \ref{fig:venues_white_literature}, and \ref{fig:type_of_web_sorces}, we reported a diverse statistical distribution of the sources used for this paper. More specifically, in Fig.~\ref{fig:type_white_literature} we have what type of studies, over the last ~15 years, have been used for our literature review. 

Experimental studies represent the prevalent type of studies found for our literature review. In experimental studies, researchers introduce an intervention (in our case, a technology), and study the effects of this intervention in a certain environment. Case studies and literature studies together make up the other half of the dataset. A case study typically entails an empirical inquiry that investigates a phenomenon within its real-life context, whilst a literature study (like this one) seeks for substantive findings and theoretical, technical, and methodological contributions to the existing body of knowledge revolving around terrorist attacks. 

 Fig~\ref{fig:venues_white_literature} graphically depicts the venues of the studies we took into consideration for this paper. Lastly, in Fig~\ref{fig:type_of_web_sorces} the two main sources of our grey literature are shown. Most of the grey literature data has been drawn from online articles, while the product pages of companies that produce technology for the protection of public spaces have less impact on the overall grey literature dataset.


\subsection{Topic Modeling}
Latent Dirichlet Allocation (LDA) has been applied to extract the most relevant themes in the textual data. Before executing the LDA topic modeling, pre-processing had to be applied to all the text. The pre-processing basically consists of (1) removing all punctuation marks and numbers; (2) standardizing terms and definitions structure-wise; (3) converting all letters to lower case; (4) removing common stop words for English grammar and syntax.

\begin{figure}
    \centering
    \includegraphics[scale=0.58]{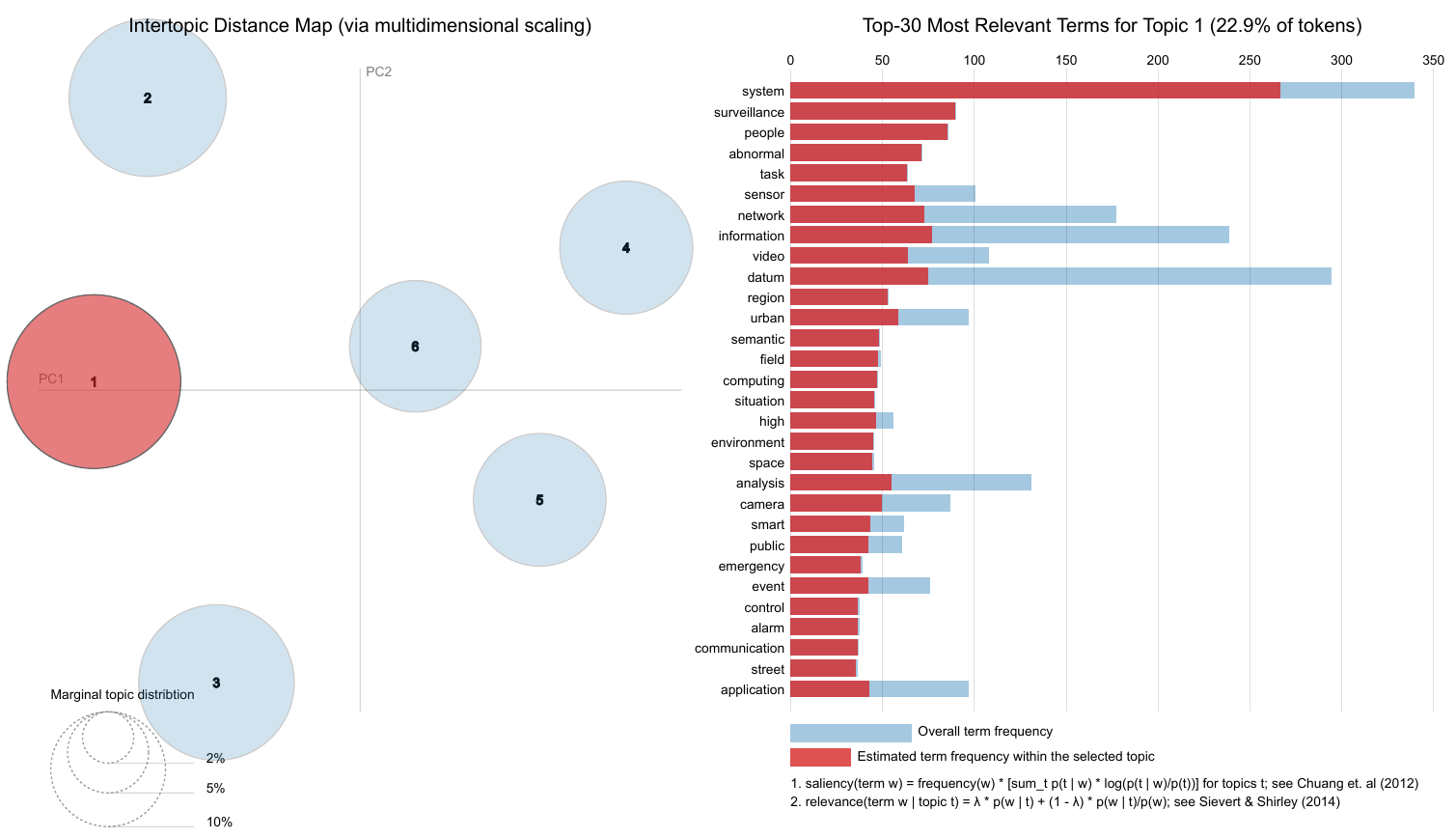}
    \caption{Topic~1 from LDA Topic Modeling. On the left side, The six circles represent the topics. On the right the most relevant terms for Topic~1.}
    \label{fig:topic1_example}
\end{figure}

After rigorous pre-processing, the LDA method for visualizing and interpreting topics was applied. The method used is called LDAvis \cite{Sievert2014} and is based on the work of Chuang, Manning, and Heer (2012). With the use of the LDAvis, visualization has been plotted, like the one in Fig.~\ref{fig:topic1_example}. Each circle represents a topic, together with its prevalence. On the right side, the top 30 most relevant terms are displayed for every topic. The $\lambda$ slider provides the opportunity for ranking the terms according to term relevance. The range of the slider goes from 0.0 to 1.0. Adjusting $\lambda$ to values close to 0 highlights potentially rare but more exclusive terms for the selected topic. Larger $\lambda$ values (closer to 1) highlight more frequently occurring terms in the document that might not be exclusive to the topic. In this research, the $\lambda$ is fixed at 0.8 to highlight frequent terms for a topic, but not exclusively. 

A bar chart renders the topic modeling results to gain more insight into the most frequent terms per topic. In this chart, we have restricted ourselves to ten terms, including their probability. The terms are ranked with their accessory probability, and those with the highest probability are displayed at the top. 

Our final research resulted in the final amount of six clusters for the white literature and two clusters for the grey literature.

For the sake of space, here below, we reported only the results of \textit{Topic~1} from the white literature to provide an example of result to the reader from LDA analysis (Fig.~\ref{fig:topic1_example}). However, in the next paragraph, we provide a taxonomy to summarize our topic modeling analysis's overall result. 

\subsubsection{Topic Modeling Analysis for White and Grey Literature}

Below we introduce the main results of topic modeling analyses. For the sake of space, we decided to depict a taxonomy overview that more concisely displays the final results.   


\begin{figure}
    \centering
    \includegraphics[width=\textwidth]{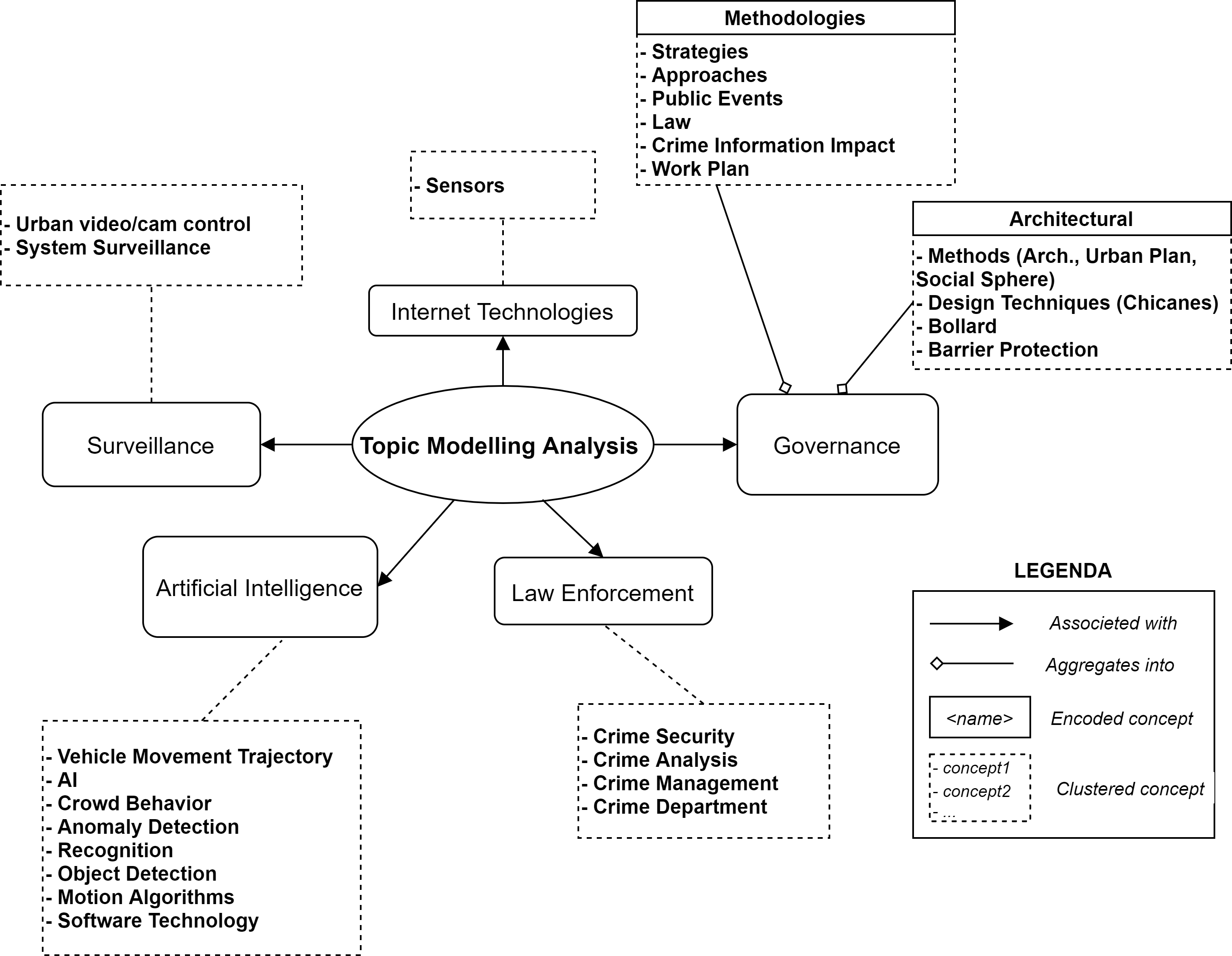}
    \caption{Topic modelling analysis taxonomy result.}
    \label{fig:taxonomy_LDA}
\end{figure}


Fig.~\ref{fig:taxonomy_LDA} depicts the results of our topic modeling analysis applied to 112 documents from academics and practitioners over the last 15 years. The \textit{Encoded Concepts} represents the six major categories found by our LDA analysis. The Encoded Concepts are a set of words that belong to a specific topic. In this group of concepts, we recognize a subset of themes defined in Table~\ref{tab:table_themes} that represents the main research topics from the analyzed studies (Internet Technologies, Governance (Methodologies and Architectural), Law Enforcement, AI, Surveillance). Differently from the Encoded Concepts, in the \textit{Clustered Concept}, we have the main branches the research is focusing on. In the following, we analyze more in detail all the \textit{Clustered Concept} we found during our analysis. Under the column \textbf{Terms}, we have the main keywords of each topic found during the analysis. Instead, under column \textbf{Score}, we have the weight-age (importance) of each keyword using \textit{lda\_model} and related to that specific topic. Hence, the higher is the score of a term, and more are the documents related to that specific research area. Lastly, the final column aims to provide a description and a context of the found terms. We decided to provide a description due to the fact that the document is meant to be used by academics, practitioners, and non-expert.

\begin{table*}[h!]
\centering
\resizebox{1\linewidth}{!}{
\begin{tabular}{p{1.5cm}p{1.5cm}p{14cm}}
\toprule
\textbf{Terms} & \textbf{Score} & \textbf{\Large Internet Technologies} \\ \midrule 
\rowcolor[HTML]{EFEFEF} 
Sensor & 60 & A Sensor is a device that measures physical input from its environment and converts it into data that can be interpreted by either a human or a machine. In the contest of protection of public spaces, Sensors can be used to improve the Surveillance Systems' accuracy rather than to be helpful in the context of transforming cities in an intelligent environment where if a sensor triggers an exception, action can be taken in order to resolve that event. Sensors can be useful to identifying drivers that run red lights or with high speed, generating predictive crime maps and risk assessment, to a real-time understanding of crowds and their behavior in urban spaces, detect aggressive behavior (Eindhoven municipality is one of the pioneers of this technology).
\\
\bottomrule
\end{tabular}
}
\caption{Topic analysis results for Sensors technologies.}
\label{tab:Sensors}
\end{table*}

Sensors from Tab.~\ref{tab:Sensors} have the value to be small and can be used in every type of scenario available in public spaces. In the context of the protection of public spaces, sensors can become extremely useful in the detection of weapons or explosives. Moreover, the strength of this technology is the of having semi-autonomous federations of sensors that share situational awareness information locally to specific areas and which operate under different conditions and parameters as others and therefore offer different insights locally, resulting in a discrete yet powerful area surveillance network.

\begin{table*}[h!]
\centering
\resizebox{1\linewidth}{!}{
\begin{tabular}{p{1.5cm}p{1.5cm}p{14cm}}
\toprule
\textbf{Terms} & \textbf{Score} & \textbf{\Large Law Enforcement} \\ \midrule 
\rowcolor[HTML]{EFEFEF} 
Crime\newline Security  & 130, 70 & Security and anti-crime policies alone are not enough for responding to all threats and vulnerabilities. Technologies like CCTV and IoT devices can help in improving the security of public spaces. However, we have a new trend from the architectural perspective so-called ``\textit{Crime Prevention through Environmental Design}'' (CPTED). The main idea behind the CPTED strategies relies on the ability to influence offender decisions that precede criminal acts. In this respect, it is important to increase the Natural Surveillance by improving the visibility of potential offenders to the general public, Natural Access Control in order to better differentiate among public and private spaces, and Natural Territorial Reinforcement by using buildings, fences, pavement, signs, lighting, and landscape to express ownership and define public, semi-public and private space.
\\
\rowcolor[HTML]{FFFFFF} 
Crime\newline Analysis & 130, 65 & To analyze and prevent crimes in urban areas, the best option available is to use the data received from the IoT devices and the surveillance cameras and CCTV in order to build a crime map of the city. From the crime, a map could then be possible to analyze the factors behind the criminal activities and then address the problems. It is hence important to understand the causes that drove a criminal activity and the location to find the right blend of solutions available in the technologies, guidelines, and architectural field.      
\\
\rowcolor[HTML]{EFEFEF} 
Crime\newline Management & 130, 60 & Police officers are the main actors in charge of crime management. Due to the police's responsibility, it is important to equip policemen with the proper technologies and grant the access to the data coming from the technological cities. Connected tablets and smartphones certified by national agencies to provide encrypted communications and in order to share documents securely and privately. Data harvested from the IoT devices by police can be analyzed to create a picture of crime patterns and trends. Leveraging on predictive analytics and machine learning to big data, police departments can foresee the place of the next violent crime. Facial recognition technology can be implemented into the CCTV and surveillance cameras and could be accessible only by police officers.
\\
\bottomrule
\end{tabular}
}
\caption{Topic analysis results for Law Enforcement Agencies.}
\label{tab:LEAs}
\end{table*}

In Tab.~\ref{tab:LEAs} we have some examples of how the LEAs can organize departments to manage, avoid and mitigate illicit behaviors in public spaces. The three departments can work together using the data retrieved from the ``technologies'' installed and disseminated in the public areas. All the data can be used and analyzed to support investigators during the investigations. Moreover, the three departments can work closely with the municipality to improve the overall life quality of those areas more exposed to possible crimes.

\begin{table*}[h!]
\centering
\resizebox{1\linewidth}{!}{
\begin{tabular}{p{1.5cm}p{1.5cm}p{14cm}}
\toprule
\textbf{Terms} & \textbf{Score} & \textbf{\Large Surveillance} \\ \midrule 
\rowcolor[HTML]{EFEFEF} 
System\newline Surveillance  & 260, 90 & Surveillance systems mostly refer to CCTV and security cameras. Moreover, in order to build a surveillance system it is necessary to include components like \emph{Cameras} the main surveillance unit, \emph{Cables} wires to connect cameras with monitors and power adapters, \emph{Power Distribution Block or Power Adapter} power supply unit, \emph{Monitors} screens for surveillance, and \emph{Video recorder (DVR)} hard drive for storage.
\\
\rowcolor[HTML]{FFFFFF} 
Urban \textbf{Video/Camera} Control & 60 & Urban Video/Camera Control systems are related to CCTV technologies. In this field, we have different types of Cameras like Analog Fixed Cameras, Analog Dome Cameras, Ip Dome Cameras, License Plate Recognition Cameras, Thermal camera (FLIR). Besides, Cameras can be used for a lot of different purposes, real-time traffic control systems in order to find anomalies, management systems for public infrastructure, identifying all anomalous events occurring on urban streets, and dealing with them in real-time, optimizing emergency, police and street service response. To achieve better results, are now available high-performance day/night cameras equipped with IR cut filter system, backlight compensation to be capable of recording into the dark, interline transfer CCD sensors.
\\
\bottomrule
\end{tabular}
}
\caption{Topic analysis results for Surveillance technologies.}
\label{tab:Surveillance}
\end{table*}

Tab.~\ref{tab:Surveillance} lists the most common technologies for the surveillance of public spaces, namely: System Surveillance and Urban Cameras. This is an adaptable technology due to the fact that nowadays, multiple models of ``cameras'' are available on the market. We can state that one of the major weaknesses of this technology is price elasticity. The range indeed can vary from a small number of euros for the IP-Cam for private and personal use to thousand of euros for those proposed by Level Five Supplies~\footnote{https://levelfivesupplies.com/} with the possibility to be mounted on a car, or the cameras proposed by Eight Bells~\footnote{https://www.8bellsresearch.com/} for accurate temperature measurement of people with atypical temperature in a crowded place, or the cameras proposed by Chris Butler Associates\footnote{https://chrisbutlerassociates.co.uk/} that integrate machine learning solutions.

\begin{table*}[h!]
\centering
\resizebox{1\linewidth}{!}{
\begin{tabular}{p{1.5cm}p{1.5cm}p{14cm}}
\toprule
\textbf{Terms} & \textbf{Score} & \textbf{\Large Governance - Methodologies} \\ \midrule 
\rowcolor[HTML]{EFEFEF} 
Approach & 65 & A broader approach is essential to guarantee safety and security in urban spaces. To this respect, it is important to have a continuous and systemic process of regulation of problems rather than their eradication or deep transformation. Moreover, it is important to know that municipalities do not have to rely only on traditional security actors, but it is vital to look broader at what are the hazards, decide who might be harmed and how to evaluate the risks, record findings, and review the risk assessment and revise if necessary \cite{1477370812473535, urban_safety}.
\\
\rowcolor[HTML]{FFFFFF} 
Work Plan\newline Strategies  & 40, 40, 40 & In \cite{actionplan1} and in \cite{actionplan2} we have some guidelines, work plan, and strategies provided by the European Commission. Moreover, a work plan can be generated using the indicators available to define Methods (those Methods from Tab.~\ref{tab:gov-architecture}).
\\
\rowcolor[HTML]{EFEFEF} 
Public Event & 18, 18 & In recent years a shift has occurred: attacks on ``lifestyle'' targets have become increasingly common. The public events always gather loads of people and the chaos of the event itself as a significant number of people, represent an easy target for terrorists that can go unnoticed. It is hence crucial to find tools, technologies, and methods that can help and assist LEAs during the crowd controls.
\\
\rowcolor[HTML]{FFFFFF} 
Law & 13 & Laws are the first ``barrier'' against crime and criminal activities. The EU is providing an action plan in this respect to support the protection of public spaces~\cite{actionplan1}. Moreover, the EU is not only providing guidance and consideration but also providing target founding, exchange of expertise and best practices, improving the cooperation among public and private sectors, and enhancing transportation security~\cite{protecting_public_spaces}.
\\
\rowcolor[HTML]{EFEFEF} 
Crime\newline Information Impact & 13, 10 & Crime information and statistics can be collected as important insights into what criminologists call the ``dark figure of crime'' or crimes that have not been reported. To assess and contrast crimes in the cities, it is important to keep track of the current crime trends and the related areas, hence, to apply strategies to contrast these crimes.
\\
\bottomrule
\end{tabular}
}
\caption{Topic analysis results for Methodologies under the Governance cluster.}
\label{tab:gov-methods}
\end{table*}

In Tab.~\ref{tab:gov-methods} are listed some approaches that could be taken by municipalities to enhance the protection of public areas. From the literature we discovered difficulties in finding approaches and methodologies to prevent, manege and avoid attacks in urban areas. From our research we could not find external companies that provide services to trainee a municipality and the related employee on how to build and reuse approaches and methodologies for the protection of public spaces.

\begin{table*}[h!]
\centering
\resizebox{1\linewidth}{!}{
\begin{tabular}{p{1.5cm}p{1.5cm}p{14cm}}
\toprule
\textbf{Terms} & \textbf{Score} & \textbf{\Large Governance - Architectural} \\ \midrule 
\rowcolor[HTML]{EFEFEF} 
Methods  & 130 & The Methods to guarantee the protection of public spaces are established on the federal and local levels. Authorities should decide and set those indicators for a positive environment concerning architecture, urban planning and social sphere \cite{RASTYAPINA20162042}.
\\
\rowcolor[HTML]{FFFFFF} 
Design & 40 & Different and various are the solutions at the design level in order to guarantee safer urban spaces. An example could be to plan streets with multiple chicanes, which require vehicles to turn corners and deliberately slow down rather than plan ways to take pedestrians off of dangerous corners but still make it convenient for them. Other examples are sidewalks lined with large trees to make it difficult for the cars to drive on, entirely pedestrianizing popular walking streets could be a solution to keep terror vehicles far from popular areas, a ring of steel to protect the most vulnerable urban areas \cite{safecity, economistreport}.
\\
\rowcolor[HTML]{EFEFEF} 
Bollard & 16 & A bollard is a short post used to create a protective or architectural perimeter. When installing primarily as a visual guide, bollards guide traffic and mark boundaries~\cite{bollard}. The bollard is only one type of protection barrier nowadays used by municipalities to protect, discourage and also as an ornament for the city to preserve and, at the same time, avoid the feeling of fear in the citizens. 
\\
\rowcolor[HTML]{FFFFFF} 
Barrier\newline Protection & 12 & It is the first approach used by municipalities to guarantee the security of urban spaces. The barriers available differ in types and purposes; an extensive list of these barriers is available on the web, an example in~\cite{barrer1,barrer3,barrer4} and some more decorative solutions like those proposed in ~\cite{barrer5}.
\\

\bottomrule
\end{tabular}
}
\caption{Topic analysis results for Architectural solutions under the Governance cluster.}
\label{tab:gov-architecture}
\end{table*}

In Tab.~\ref{tab:gov-architecture} we have some architectural approaches. Together with the surveillance systems, barriers and bollards proved to be the most used approaches. However, misuse of this type of approach to securing public spaces could be seen by many as `disproportionate'. Moreover, it profoundly impacts the liveability, walkability, character, and accessibility of public spaces. Security, in this sense, does not provide feelings of safety and security and, indeed, can have the opposite effect. However, nowadays, the predominant emerging view is to try to be camouflaged and subtly embedded within the cityscape the barriers to protect public spaces. It is hence strategic to rethink barriers to be appropriately integrated into the cityscape. However, the idea of embedding security measures into the design plans of our cities and public spaces is becoming more and more predominant.

\begin{table*}[h!]
\centering
\resizebox{1\linewidth}{!}{
\begin{tabular}{p{1.5cm}p{1.5cm}p{14cm}}
\toprule
\textbf{Terms} & \textbf{Score} & \textbf{\Large Artificial Intelligence} \\ \midrule 
\rowcolor[HTML]{EFEFEF} 
Motion\newline Algorithms  & 75, 50 & Conventional video surveillance systems often rely on human operators for activity monitoring and determining actions to be taken upon incident occurrence. The new era of video surveillance systems provides support to the tedious work of human operators obliged to watch recorded videos in order to find crime proofs. The industry and academics have developed technologies for intelligent surveillance, such as object tracking \cite{4042701,5475233}, pedestrian detection \cite{Dalal:2005:HOG:1068507.1069007}, gait analysis \cite{Wang:2006:AWG:1170749.1172660},
vehicle recognition \cite{4202577}, face and iris recognition \cite{5471147}, and crowd counting \cite{5206648}.  
\\
\rowcolor[HTML]{FFFFFF} 
Object\newline Detection & 60, 90 &  We can state that for object detection we can use the same technology presented in the other tables from the other topics (motion algorithms, AI, IoT devices).
\\
\rowcolor[HTML]{EFEFEF} 
Crowd\newline Behaviour  &  50, 70 & Crowd Behaviour can be easily addressed by artificial intelligence technology implemented in CCTV and urban cameras. In order to have a broad overview of the city is possible to implement a network composed of AI cameras, IoT devices, and analysis systems so to have a picture of the security state of any urban areas analyzing the data collected from all the different devices at runtime. Moreover, AI itself is already able to detect crowd behavior and anomalies and can be implemented in drones and CCTV, an example of this technology nowadays available are described in \cite{10.1007/978-3-319-59147-6_26,verge1,Singh2018EyeIT,verge2} and are available in real-time webpage\footnote{https://store.icrealtime.com/cameras} and in Boulderai webpage\footnote{https://www.boulderai.com/our-hardware/}.
\\
\rowcolor[HTML]{FFFFFF} 
 Recognition & 50 &  Latest CCTV are equipped with AI in order to recognize different types of objects and scenarios. Hikvision, one of the main suppliers of video surveillance products and solutions, in 2017, launched the first 'Deep Learning' embedded in a Network Video Recorder (NVR) \cite{hikvision}. The technology presented by Hickvision performs different types of recognitions: \emph{Facial Recognition}, \emph{People counting}, and \emph{Management of vehicles}. The same technology has also been implemented by Huawei and Gigabyte companies, and the related products are already available for the market directly from the Huawei webpage\footnote{https://e.huawei.com/en/products/intelligent-video-surveillance/cameras/software-defined-camera/x3221-c} and the Gigabyte store\footnote{https://www.gigabyte.com/Solutions/AI-AIoT/intelligent-video}.
\\
\rowcolor[HTML]{EFEFEF} 
Anomaly\newline Detection &  40 & As for Crowd Behavior, we can use same methods and techniques described also for the detection of anomalies. 
\\
\rowcolor[HTML]{FFFFFF} 
Vehicle Movement Trajectory & 30, 30, 30 & The CCTV and video cameras are nowadays extensively used in order to guarantee safety and security in public spaces. However, in \cite{tno}, TNO developed WAMI (Wide Area Motion Imagery), where drones are equipped with high-resolution cameras in order to observe an entire city continuously from above. The algorithm proposed by TNO includes object detection, object tracking, tracking repair, and track analytics. WAMI can be used by the analyst to study the behavior of people and vehicles. In \cite{Mehboob2019} they propose a video visualization system for traffic surveillance. Based on the glyph, the tool can be utilized for road surveillance videos to monitor live road traffic on the highways.
\\
\rowcolor[HTML]{EFEFEF} 
Software Technology  & 17, 18 & The progress of the technology for the protection of public spaces are visible to everyone. In some cases, the software is even considered a threat and too intrusive of personal privacy. An example is the case of facial recognition in the UK~\cite{facialrecognition}. Hence, it is necessary for the developers, LEAs, and municipalities to find a good trade-off between the protection of urban areas and the right software and technology to employ to do not undermine the privacy of the users. 
\\
\rowcolor[HTML]{FFFFFF} 
AI & 5 &  Artificial Intelligence is the latest and most promising silver bullet proposed by academia for the protection of urban areas. An increasingly sophisticated technology, AI could support preventive policing to bring about a safer community. The technology is still at an early stage, and a lot of work is needed to improve it. However, the results of AI are promising and of a wide range of appliances.  
\\
\bottomrule
\end{tabular}
}
\caption{Topic analysis results for Artificial Intelligence technologies.}
\label{tab:artificial_intelligence}
\end{table*}

Lastly, but not least, in Tab.~\ref{tab:artificial_intelligence} are listed some upcoming approaches using Artificial Intelligence (AI) and Machine Learning (ML) algorithms to provide protection of public spaces. These type of approaches can be directly integrated in surveillance cameras rather then can be used to automatically monitoring the data coming from different sources.

\subsection{Topological Analysis}
\label{topanal}
After the Topic Modelling analysis on the whole dataset containing both white and grey literature, we have applied Topological Data Analysis (TDA) to exploit the topological and underlying geometric structures of our data. As a data science methodology approach, TDA seeks to identify patterns in big data, usually to generate predictors. 
TDA attempts to grasp a dataset at a more meaningful level. We apply TDA to find similarities in our dataset and to eliminate all those data that are not relevant for the analysis to find patterns and shapes in our data. However, TDA's main idea is to focus on what makes data points similar to each other (e.g., row-by-row comparisons) and using this to compress the information into an interactive graphical representation of the aggregate data. TDA looks at the at the shape of our data. With TDA analysis we aim at intercepting anomalies or outliers that have not been highlighted in the topic modelling analysis. This graphic representation depicts groups, many previously unseen, that naturally occur in the data~\cite{TDA}. Below, we have the graphical representation of our dataset. 
\begin{figure}[h!]
    \centering
    \includegraphics[width=0.7\linewidth]{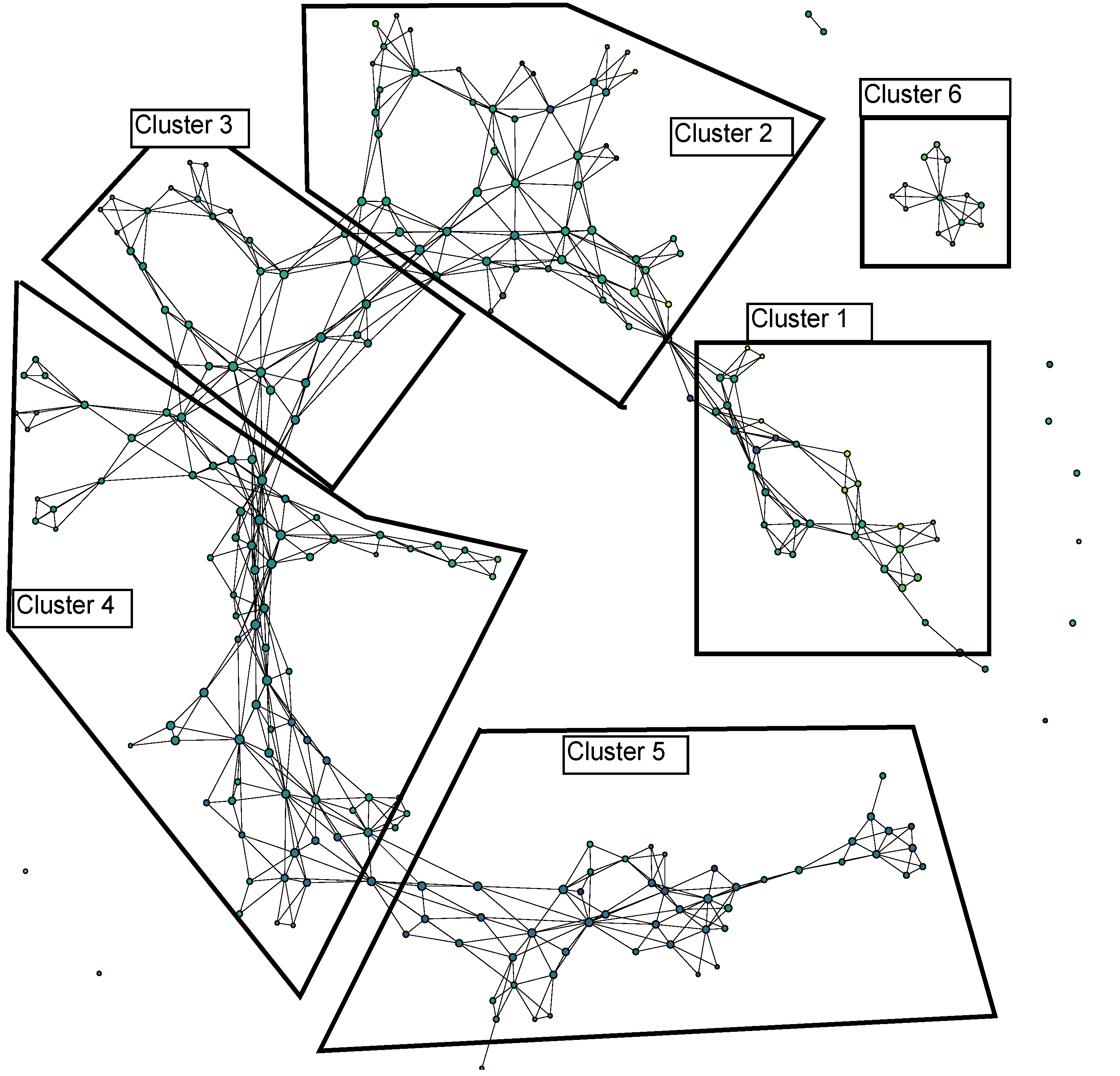}
    \caption{Topological Data Analysis of the data from the white and grey literature.}
    \label{fig:TDA_analisys}
\end{figure}

The output in Fig.~\ref{fig:TDA_analisys} represent the shape of our dataset, and as depicted in the picture, we have six significant datasets.

\begin{table*}[h!]
\centering
\scalebox{0.75}{
\begin{tabular}{p{0.24\textwidth}p{1.0\textwidth}}
\toprule
\textbf{Cluster} & \textbf{\Large Cluster Detail} \\ \midrule 
\rowcolor[HTML]{EFEFEF} 
Cluster~1  & Security measures, Personnel, Prevent, Drones alert, Law, Agents    understand terrorists attacks, Pattern recognition.  
\\
\rowcolor[HTML]{FFFFFF} 
Cluster~2 & Artificial Intelligence, Vehicle, Public safety, Traffic management,
Recognition, Drones, Smart city, Remote technology, Decision making, Neural network.
\\
\rowcolor[HTML]{EFEFEF} 
Cluster~3 & Critical perimeter, Detectors, Build simple detectors, Intelligent Communications emergency, Urban infrastructure, Sensors detect persons, RFID.
\\
\rowcolor[HTML]{FFFFFF} 
Cluster~4 &  Awareness, Simulation, Physical layer, Architecture, Physical resources, Crime patterns, Video surveillance, Crime strategy planning.
\\
\rowcolor[HTML]{EFEFEF} 
Cluster~5 & Trajectories, Anomalous motion patterns, Abnormal behaviour, Anomaly detection.
\\
\rowcolor[HTML]{FFFFFF} 
Cluster~6 & Fog Computing, Smart city, Cloud, IoT.
\\
\bottomrule
\end{tabular}
}
\caption{Cluster details of Topological Data Analysis.}
\label{tab:TDA}
\end{table*}

It is important to underline that the LDA (topic modeling) analysis is not comparable with the TDA since the two analyses have different aims and approaches. LDA discovers thematic features and their respective structures in a
large number of studies using a probabilistic topic model. While TDA is a methodology to study the topological shape of data. The fundamental assumption in topology is that connectivity is more important than distance. Hence, with TDA, we aim to analyze relationships between a set of documents and the terms they contain.

From Table~\ref{tab:TDA} we have been able to extract six most representative clusters. From Cluster~1 and Cluster~4 we infer information related to LEAs and Governments approaches. Indeed, we recognize terms like \textit{security measures, law, agents understand terrorist attacks, architecture, physical resources} that give us information about those methodologies implemented by the municipalities and LEAs. In Cluster~2, Cluster~3, and Cluster~5 defines all those technologies, algorithms, and cyber-physical systems that are able to convert a municipality into a Smart City. Lastly, Cluster~6 encompasses all those technologies related to the IoT. The fog computing extends the concept of cloud computing to the network edge, making it ideal for internet of things (IoT) and other applications that require real-time interactions. As mentioned previously, we applied TDA analysis to find outliers in our dataset and to catch information that had not been captured with the topic modeling. Overall we can assert that the information found with the topic modeling analysis finds confirmation in the TDA analysis. Clusters from one to five from TDA contain information highlighted with the topic modeling analysis. However, Cluster~6 appears to add some new technologies for the protection of public spaces like \textit{Fog Computing} and \textit{Cloud} infrastructure that have not been captured with LDA.

\subsection{t-Distributed Stochastic Neighbor Embedding (t-SNE) Analysis}

To further analyze our dataset, we decided to apply a t-Distributed Stochastic Neighbor Embedding analysis. t-SNE is an unsupervised, non-linear technique for dimensionality reduction that is particularly well suited for the visualization of high-dimensional datasets. The t-SNE algorithm calculates a similarity measure between pairs of instances in the high dimensional and low-dimensional space. It then tries to optimize these two similarity measures using a cost function. 
\begin{itemize}
    \item Step 1: in the high-dimensional space, create a probability distribution that dictates the relationships between various neighboring points.
    \item Step 2: It then tries to recreate a low dimensional space that follows that probability distribution as best as possible.
    \item Step 3: We measure the difference between the probability distributions of the two-dimensional spaces using Kullback-Liebler divergence (KL)~\cite{t-SNE1,t-SNE2,t-SNE3}.
\end{itemize}

We decided to apply t-SNE analysis to further understanding of how the data is arranged in a high-dimensional space. 

\begin{figure}[h!]
    \centering
    \includegraphics[width=0.75\linewidth]{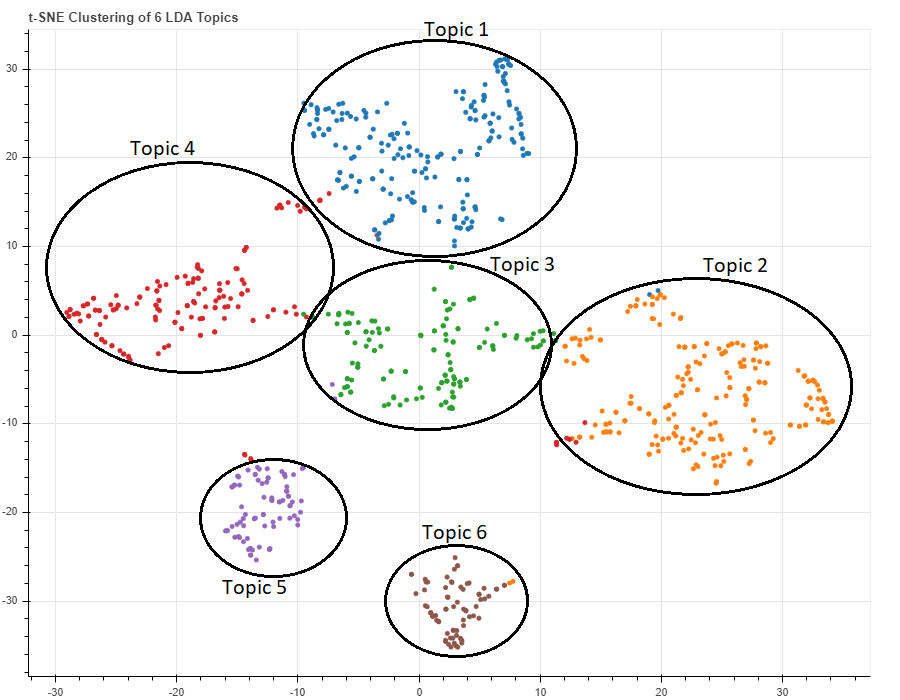}
    \caption{t-SNE analysis of the data from the white and grey literature.
    }
    \label{fig:t-SNE_analisys}
\end{figure}

Clearly Fig.~\ref{fig:t-SNE_analisys} exhibits distribution of the six topics from our LDA topic analysis. To gain a deeper insight of the six topics we provided a word cloud image where the most representative words of each topic are listed. 


\begin{figure}[h!]
    \centering
    \includegraphics[width=0.65\linewidth]{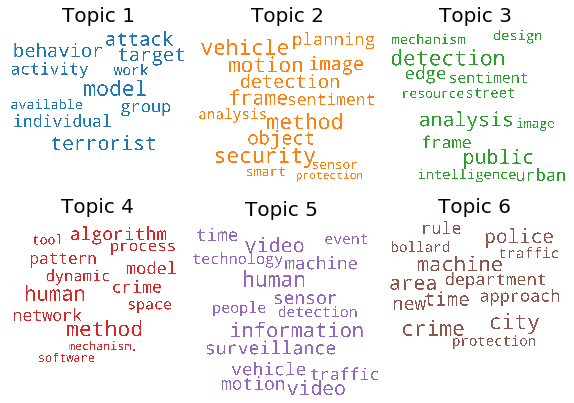}
    \caption{Word Cloud with the most representative words per each topic.}
    \label{fig:word_cloud}
\end{figure}

From the six word cloud in Fig.~\ref{fig:word_cloud} we see the distribution of the six topics. Fig.~\ref{fig:word_cloud} is the textual representation of the clusters from the analysis available in Fig.~\ref{fig:t-SNE_analisys}. We can group Topic~5, and Topic~6 under the \textbf{Governance} and \textbf{Law Enforcement} umbrella. While, Topic~1, Topic~2, Topic~4, and Topic~5 represent the cyber physical systems that can be organized under \textbf{Law Enforcement}, \textbf{Artificial Intelligent}, \textbf{Surveillance}, and \textbf{Internet Technologies}.
Moreover, from the analysis in Fig.~\ref{fig:t-SNE_analisys} we infer how Topics from 1 to 4 are stochastically close, while Topic~5 and Topic~6 result more disconnected compared to the other clusters. This could be interpreted with the fact that in Topic~5 we have information strictly related to IoT devices while in Topic~6 we find documents related to techniques, best practices, and approaches for LEAs, municipalities, and governance. Conversely, in the rest of the documents, we have the most information about CPS like motion tracking, deep learning, and face recognition.


To conclude, we decided to apply t-SNE analysis to investigate the distribution of our data further, to (a) find conceptual relations across clusters, (b) confirm the results found with our Topic Modelling analysis, and (c) lastly, to give a better understanding of our findings and our research.   

\subsection{Distribution of theme codes in White and Grey literature.}

\begin{figure}
\begin{minipage}{.56\linewidth}
\centering
\subfloat[]{\label{main:a}\includegraphics[scale=.40]{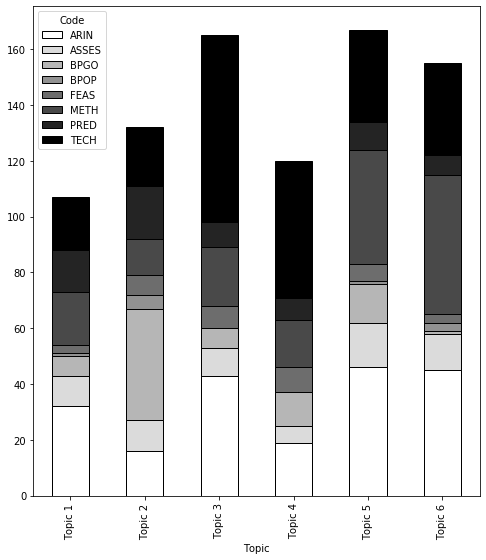}}
\end{minipage}%
\begin{minipage}{.5\linewidth}
\centering
\subfloat[]{\label{main:b}\includegraphics[scale=.38]{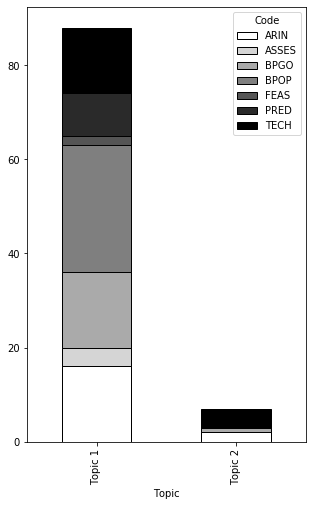}}
\end{minipage}\par\medskip

\caption{Amount of codes present in the white (a) and grey literature (b).}
\label{fig:amount_of_codes_white_grey}
\end{figure}

To conclude the discussion of our analysis, we provide in Fig.~\ref{fig:amount_of_codes_white_grey} the distribution of all the themes presented in table~\ref{tab:table_themes} in our dataset composed of both white and grey literature. In Fig.~\ref{fig:amount_of_codes_white_grey}, (a) refers to the distribution of the documents from the white literature. Meanwhile, in Fig.~\ref{fig:amount_of_codes_white_grey} (b), we have the documents related to the grey literature, with Topic~1 as a dominant cluster and Topic~2 as smaller one.  
In Fig.~\ref{fig:amount_of_codes_white_grey} (a), we can observe that the scientific literature covers almost all the fields codified in table~\ref{tab:table_themes}, with a higher amount of documents related to Artificial Intelligente (ARIN), Technologies (TECH), and Best practices (BPGO - BPOP). Conversely, in Fig.~\ref{fig:amount_of_codes_white_grey} (b), even if the distribution of the grey literature appears to be covering almost all the codes from table~\ref{tab:table_themes}, we have a good amount of Best practices (BPGO - BPOP), but a relatively small amount of documents related to Artificial Intelligence (ARIN) and CPS Technologies (TECH). The low number of CPS technologies is due to the fact that most companies that are providing this type of technology do not provide enough information online. Most of the time, the websites of these companies provide an online brochure or just a simple overview of the services and technologies they provide. Hence, it has turned out exceeding difficult to detect them using structured queries.

%% file: sections/5_discussions.tex
\section{Discussions}
\label{discussions}

A major point of discussion for the proposed work, is that a concrete overlap exists among the defined codes and the topic modeling results. Indeed, Table~\ref{tab:Sensors} and Table~\ref{tab:Surveillance} present a list of terms associated with cyber physical systems (\textbf{TECH}) used in urban spaces to guarantee safety and security. In Table~\ref{tab:LEAs}, we enumerated techniques to assess and prevent (\textbf{ASSES}) crimes,  which range from how to prevent crimes and how to analyze those urban areas more subject to criminal events to how to deal and manage areas where criminal events are more frequent. In Table~\ref{tab:gov-methods} and in Table~\ref{tab:gov-architecture} our topic modeling analysis grouped the best practices (\textbf{BPGO, BPOP}) that can be used both at municipal or government levels in order to improve the safety and security of urban spaces. This topic groups methods, approaches, and new architectural design techniques in order to improve the environment at the structural engineering level, a way to rethink our cities and design new spaces and new urban areas that can be considered secure against criminal activities by design. Continuing with our topic modeling analysis results, in Table~\ref{tab:artificial_intelligence} are presented cyber physical system that leverage Artificial Intelligence algorithms (\textbf{ARIN}) in order to enhance the capacities of CCTV, drones, and surveillance cameras. 
Considering our overall analysis, the main research question can be answered:\\
\par
{\it ``What are the best practices and CPS technologies, nowadays available, for the protection of urban areas?''}
\begin{framed}
        \textbf{Major Findings.} There is no one single bullet-proof solution available. Conversely, multiple methods and techniques can be put in place to guarantee safety and security in public spaces. The techniques range from architectural design to rethink public spaces keeping security into account in continuity to emerging technologies such as AI and predictive surveillance. Among the architectural field the barriers hidden into the landscape of the city appear to be a good option for protecting public spaces with a relatively low maintenance cost. On the other side ML and AI implemented in CCTV emerged to be an incisive technology while extremely costly. However, whenever new technologies can appear expensive in cost and questionable in privacy or could be challenging to rethink and re-project public spaces, it is always possible to build an action plan to mitigate, prevent and manage crime events. 
\end{framed}

Furthermore, the sub-questions can be addressed as follows:\newline

\textit{(SRQ1.) To what extent new technologies effectively enhance the protection of public spaces?}
This research question was originally aimed to assess what tools are available for the protection of public spaces. Our data indicate an array of technologies spanning from the Artificial Intelligence, Machine Learning, CCTV that are already implementing these technologies for the recognition of human faces and traffic analysis.
\begin{framed}
    \textbf{Finding~1.} From our analysis we found three major active fields for the protection of public spaces. The \textbf{Information Technology} (IT) field offers different innovations like modern cameras with integration of AI, thermal cameras, sensors, machine learning models for object detection and crowd behaviour. However, on the other side the IT field offers cheaper technologies like automated bollards and barrier protection. From the analysis we have intelligent barriers able to be activated through sensors and be an active part of a smart city. 
    The second field is the \textbf{Architectural} one. Under the architectural field our analysis showed the possibility to redesign the public space to make them safe and protect citizens. Moreover, the architectural field showed the possibility to integrate static barriers into the city landscape reducing the sense of fear. Lastly we have the \textbf{Organizational} field. Under the organizational field our analysis showed some major approaches both for municipalities and LEAs. Approaches like \textit{work plan} and \textit{work plan strategies} during public event can be designed and organized directly by the municipalities in conjunction with LEAs, fire fighters and healthcare. On the other side LEAs can work on plans for \textit{crime analysis}, \textit{crime management}, and \textit{crime security} working with data scientist and collaborating with academic to study and set up new approaches.

\end{framed}

\textit{(SRQ2.) What are the new approaches for the protection of urban areas?}
This research question was initially aimed to assess what CPS and best practices are available for the protection of public spaces. Our data highlighted a wide range of new techniques and best practices for the protection of urban areas, i.e., new architectural approaches in order to have spaces and buildings thought to protect people, crime and statistics studies in order to predict the most sensible areas in a city, EU approaches and techniques to help municipalities and LEAs to enhance the security levels of the cities.\newline    

\begin{framed}
    \textbf{Finding~2.} 
    Our analysis highlighted how both academics and the market are pursuing a model of the Smart City~2.0. An always more smart city and a smart environment is the next challenge for both academic researchers and practitioners. Indeed the usage of CPS like sensors, machine learning, and artificial intelligence, resulted in being predominant in the overall SLR. We hence can reply to \textit{(SRQ2.)}, that the new approaches for the protection of public spaces are smart technologies, able to provide a huge amount of data about the ``health'' of the smart city, while demanding a less human interaction, hence providing a high grade of security for all the citizens.   
\end{framed}

\textit{(SRQ3.) To what extent is the agreement between white and grey literature, and what are the gaps?}
This research question was initially aimed to assess the gaps and the overlap between academia and the research produced by organizations outside of traditional academic publishing. The purpose of \textit{(SRQ3.)} is hence to provide new research opportunities or new cues of reflection to the academic community.

\begin{framed}
    \textbf{Finding~3.} 
    To answer \textit{(SRQ3.)} we use a set theory based graphical representation of two concentric circles, where the outermost circle represents the academic, and the internal one is the market. In the field of CPS for the protection of public spaces, most of the research produced by academia is converted into products by the market. Moreover, is necessary to highlight that if from one side the academia, in the context of CPS, implements most of the algorithms, on the other side the market is the one responsible for the hardware and the final implementation, hence, in this respect, the two actors are complementary each other.   
\end{framed}
    

Overall, on the one hand, we report a consistent overlap between areas of research in CPS technologies for the objectives in our vision, but, on the other hand, we cannot assert the same for what concerns best practices and methodologies for the protection of public spaces. Indeed we found that:

\begin{framed}
    \textbf{Finding 4.} While the academics suggest best practices and methodologies for the protection of urban areas, the market did not provide any type of implementation of such suggested approaches, which shows a lack of fertilization between academia and industry. A relevant example is given by the European Community that in the context of protection of urban spaces provides guidelines and methodologies through the EU portal and available for all the EU cities. 
\end{framed}  

Furthermore, the best practices identified in websites and blogs are most of the time, reports from the European Community, which highlight a community imbalance in terms of global distribution of research efforts; this highlights a still very much primordial field of research wherefore the foundations we provide may help to bring order and more structured research roadmaps all across the globe. 

Interestingly, we observed the same imbalance also in terms of the architectural approaches: on one side, we report the academics that are suggesting new architectural approaches for safety and security-oriented smart buildings, while the market and practitioners often do not invest and implement/adopt new concepts. 

In this regard, the scientific community must start producing methodologies, guidelines, and architectural solutions that municipalities can easily implement to enhance security in public spaces. The production of such guidelines and methodologies of easy use and applicability from the academic needs to have two main objectives: 1) not to let the cities be dependent only to technologies, 2) reduce deployment and general implementation/operation costs, 3) improve collaboration among municipalities, LEAs, fire-fighters, health-care systems, and all stakeholders involved in the risk assessment, mitigation, and management of crises and cyber-physical threat events such as terrorist attacks.

%% file: sections/6_lesson_learned.tex
\section{Lesson Learned}
\label{lessonlearned}

From Section~\ref{discussions} we want to extract our ``lesson learned'' and some useful guidelines to be used by practitioners to develop new algorithms to detect illicit behaviors, for LEAs and municipalities to improve safety and security in public areas, and for academics to better target future research directions.
As already stated previously, we could not find a definitive solution to guarantee a higher level of security and protection of urban spaces against terroristic attacks, illicit behaviors, illegal actions, and, more in general criminal acts and activities. Conversely, we demonstrated how the online literature provides multiple approaches to cope same problems in public areas.

\begin{itemize}
    \item Practitioners mainly focused on tools and technologies. There is a lack of practitioners providing services to train municipalities and LEAs in building and implementing best practices for the protection of public spaces. 
    \item LEAs and Municipalities could introduce the usage of data analysis to monitor the city and map the crime areas. Moreover, stricter collaboration could improve overall security in public spaces.
    \item Academia showed a nice overlap with practitioners regarding new technologies and algorithms. However, we notice a bad interaction among multiple research fields like architecture and information technology in order to integrate technologies within the new buildings. 
\end{itemize}

In the following, we provide a more extensive explanation of our ``lesson learned'' together with motivation and an explanation of our findings.

\emph{Practitioners}: represent the business and governmental part of this systematic literature review. Our research revealed that the majority of commercial-strength solutions and tools had been delivered by security software and tech industry instead of R\&B organizations. Conversely, it seems that the business side of our review is lacking in providing services to build and implement best practices based on the different cities and different sets of problems to solve. Furthermore, most of the time, the main provider of best practices appeared to be the European Community, with a set of new standards, guidelines, and approaches to fight terrorism attacks in public spaces. In this regard, based on our SLR we firmly believe that a serious challenge is to encourage tech providers to propose new businesses where best practices support the technologies. More in detail, we aim to inspire tech providers and practitioners in providing not only technologies but also best practices regarding the usage of these technologies and the appropriate environment and scenario where to install and use them.

\emph{LEAs - Municipalities}: both of them represent the main technology consumers. LEAs and municipalities rely on technologies and tech providers to, from one side, speed-up investigation using more reliable data and to prevent crimes, while, on the other side, municipalities are looking to safer places to bring together the community and to organize secure events. From our study, we notice that both these actors are using technologies against crime. However, it is important to underline how the two actors should improve the collaboration (maybe by sharing more data) to fight crime. Indeed, it is still rare the usage of data regarding crimes to build a city map to highlight the zones with the highest rate of perpetrated crimes. In our opinion, the introduction of city zones with elevated crime risk could be, from LEAs side a solution to mitigate and prevent disorders and illicit behaviors, while from the municipality's point of view could be a reason to invest effort and money to improve the overall life quality in the risky zones. However, city mapping it is not the only solution that arise from our literature review. Indeed, undertaking a study of the geographical area of the municipality can lead to implementing more sophisticated architectural buildings, squares, and parks that can improve the security of the urban spaces. Lastly, and probably the most important lack in the protection of public spaces that arise from our study is the lack of sharing best practices among different LEAs and different municipalities. It is of vital importance to build a shared repository at the European level, where LEAs and municipalities can provide best practices documentation applied for specific scenarios and in a specific environment. This type of shared best practices can help cities with lower technological resources and lower budgets to improve the overall quality of safety and security in their public spaces. Lastly, from our SLR, we can state that the municipalities have two significant approaches to improving the city's overall security. The first approach, we called it ``economically-wise''. Municipalities can decide to invest money in a more sophisticated technology like CCTV with AI models for recognizing illicit behavior. Otherwise, they can decide to spend considerably less money choosing barriers that can be hidden in the city. On the other side, we have the ``time-wise'' approach. Here the municipalities can decide to invest more time in changing the aspect of the city architecturally. In this regard, municipalities can decide to implement new types of buildings, squares, and city backgrounds thought to protect humans. On the contrary, municipalities unwilling to invest time can implement best practices.

\emph{Academics}: together with part of the practitioners represent the principal investigators of new algorithms and approaches to implement. As already discussed in our previous sections, our literature review highlighted an interesting overlap between the new technologies developed in academia and the subsequent technologies available from the markets. However, we noticed a small interaction among multiple academic fields and areas of expertise in this regard. Our direction for academics is the one to improve the overall interaction among the different fields of expertise. Hence, it is of vital importance to create a team composed of Information Technology professionals to work on new technical approaches, architects to rethink the cities more safely, and psychologists to safeguard and protect the livability, walkability, character, and accessibility of public spaces trying to avoid any sense of fear in citizens.

%% file: sections/7_limitations.tex
\section{Limitations and Threats to Validity}
\label{limitations}

Several scopes and applicability limitations of our results emerged during our study, as well as several threats to the validity of our results. This section provides an outline of both.

First, technologies such as face-recognition technology, event prediction based on human-observation data, crowd behavior, vehicle movement trajectory recognition, and more in general, AI-based analysis and synthesis information systems are predominant in our sample but pose a considerable concern about personal privacy. Threats to privacy revealed by our analysis include: \emph{Lack of Transparency} where people are not aware or did not give the consent to the collection of personal data in public spaces; \emph{Misuse} the images retrieved can be used for different purposes; \emph{Accuracy} could happen to have false-positive matches; human-centered \emph{Automated Decision making} where technology can influence the decision of investigators. The debate about the use of AI in public spaces is in its beginnings and may be the most predominant factor in correlation with the disruptive improvements of AI technology. Our systematic literature review presents the possibility to re-think and re-design AI-based systems concerning public spaces, their restrictions, and features from the architectural level to increase safety and security in public spaces. Lastly, from our SLR, we infer the possibility of building best practices and outlining action plans to prevent, manage, and analyze crime scenarios and terrorist events. 

Concerning the above approaches, this SLR does not strive for a final silver-bullet solution. Conversely, our work gives an overview of the different approaches available in the scientific and non-scientific online literature to help municipalities, governmental, and non-governmental organizations define the best approach possible for the protection of urban areas against crimes and terroristic attacks.  
  
 Based on the taxonomy in Wohlin et al. \cite{wohlin}, there are four potential validity threat areas, namely: external, construct, internal, and conclusion validity. 

\emph{ ``External Validity''} concerns the applicability of the results in a more general context. Since our primary studies are obtained from a large extent of disciplines, our results and observations might be only partially applicable to cyber-physical systems for the protection of public spaces disciplines, this may threaten external validity. To strengthen external-validity, we leveraged information from workshops held by the five municipalities and LEAs involved in the project. We analyzed this qualitative data to better overview CPS and, more in general, fine-tune our research methods and the applicability of our results.

\emph{``Construct Validity''} and \emph{``Internal Validity''} concern the generalizability of the constructs under study, as well as the methods used to study and analyze data (e.g., the types of bias involved). To mitigate these threats, we adopted a mixed-methods research approach. On the one hand, a formal grounded-theory method was conceived to avoid bias by construction~\cite{straussian,gt,gtmeth}. On the other hand, we adopted machine-learning and topic modeling techniques, which were appropriately fine-tuned using state of the art approaches to the purpose of elaborating the expected results. As previously explained, to ensure internal and construct validity even further, the initial set of codes for grounded-theory has been based on the one of Popp et al. In addition, we applied grounded-theory in two rounds: (a) first, the primary studies were split across a group of students to apply grounded theory; (b) in the second round, one of the authors re-executed a blind grounded-theory on the full primary studies set. When both rounds were finished, both grounded-theories were analyzed evenly to construct a unique theory. When a disagreement between the two samples was found, a session was organized with students, researchers and supervisors to examine the samples and check them against literature.
Some of our results are subject to discriminatory exploitation from a social policy perspective. I.e., the idea we introduce city safe zone might be discriminatory with respect to those or other zones, a sampling of our results and recommendations through an ethics advisory board is currently ongoing and will be destined to further publication. It is beyond the scope of this paper—whose objective is again to lay down a solid overview of information and software technologies for cyber-threat intelligence over urban spaces—to comment and contextualize such findings as the reviewer would seem to imply. Indeed, the discriminatory nature pointed out by the reviewer was one of the several aspects evaluated in the controlled experimentation performed as a follow-up exploitation of our findings and is under the care and publication of other units involved in the same project.

\emph{``Conclusion Validity''} concerns the degree to which our conclusions are reasonable based on our data. The logical reasoning behind our conclusions is dictated by sound analysis of the data through grounded theory and other analysis methods that try and construct theory from data rather than confirming initial hypotheses, as explained in \cite{gtmeth,gtval}.

%% file: sections/8_conclusions-future_work.tex
\section{Conclusion and Future work}
\label{conclusions}

The systematic literature review on best practices and CPS technologies for the protection of public and urban spaces presented here aims at building a comprehensive overview of what are the new trends for the protection of public spaces against terrorist attacks and in order to detect anomalies, prevent crimes, and generally how to build safer cities for citizens. First, papers have been collected from different online sources, both from the academic field and from websites and blogs. Once the data set was completely set up and organized, we began with the coding phase. The codes used are listed in Table~\ref{tab:table_themes}. Once all documents from the data set have been categorized, LDA Topic Modelling has been applied to the white and grey data set to extract the most relevant themes and compare the results with the codes used for our categorization.
Specifically, our LDA Topic Modelling analysis encoded five major concepts depicted in the taxonomy from Fig.~\ref{fig:taxonomy_LDA} and namely: \textit{Internet Technologies}, \textit{Governance}, \textit{Law Enforcement}, \textit{Artificial Intelligence}, \textit{Surveillance}. These major concepts represent the five relevant categories of CPS technologies for the protection of urban and public spaces. To further analyze our dataset, we applied Topological Analysis to identify patterns and depict the shape of our data. Finally, the t-SNE analysis provided a distribution of the six topics from our LDA topic analysis. Among all the analysis we highlighted the overlap of the results as well as all the different aspects of the analysis. Hence, we can state that it exists a concrete overlap between the defined codes and the topic modeling results. 
The overall study and analysis lead us to conclude that does not exist one single solution for the protection of urban spaces; meanwhile, multiple approaches and different CPS technologies can be put in place to avoid, mitigate, and handle terrorist threats. 

In the future, we plan to: (a) implement a proof-of-concept of recommended system to provide for online browsing of the proposed results; (b) conduct field studies of the technologies found by means of our municipality partners, to find a stable \emph{kit} or set of technologies to address their shortcomings thus evaluating technological effectiveness in simulated scenarios; (c) provide for augmented-reality facilities to implement those cyber-physical technologies which are matching municipality requirements emerged in our workshops but are still currently unknown and/or under-experimented, this includes boundary-trajectory mining, real-time violence detection, etc.; (d) elaborate a more instrumented and smart online geo-fencing system \cite{PongpaichetSJP13} that implements parts of the results of this survey with the goal of supporting municipalities in decision-making during smart-events.

%% file: sections/9_annex_I.tex
\newpage

\section*{Annex I}
In Table~\ref{tab:terms_table_pop}, key terms and related definitions used in this study are mentioned. The table consists of two columns: \textit{Terms}, where the terms which are less commonly known and more technical are mentioned. In the second column \textit{Definitions}, a short explanation is given to help better understand the essence of the mentioned term.

\begin{footnotesize}
\begin{longtable}{p{0.21\textwidth}p{0.73\textwidth}}
\toprule
\textbf{Terms} & \textbf{Definition} \\ \midrule
\rowcolor[HTML]{EFEFEF} 
Biometrics & \begin{tabular}{p{0.72\textwidth}}Identify and or verify human terrorist (or watchlist) subjects using 2D   and 3D modeling approaches over a variety of biometric signatures: face, gait, iris, fingerprint, voice. Also exploit multiple sensor modalities: EO, IR, radar, hyper-spectral \cite{popp2004countering}\end{tabular} \\
Clustering & \begin{tabular}{p{0.72\textwidth}}Employ numerous technical approaches (natural language processing, AI, machine learning, pattern recognition, statistical analysis, probabilistic techniqes) to automatically extract meaning and key concepts from (un)structured data and categorize via an information model (taxonomy, ontology). Cluster documents with similar contents \cite{popp2004countering}.\end{tabular} \\
\rowcolor[HTML]{EFEFEF} 
Event detection & \begin{tabular}{p{0.72\textwidth}}Monitor simple and complex events and notify users (or applications) in real-time of their detection. Monitoring can be scheduled a priori, or placed on an ad hoc basis driven by user demands. When an event is detected, automatic notifications can range from simple actions (sending an alert, page, or email) to more complex ones (feeding information into an analytics system) \cite{popp2004countering}.\end{tabular} \\
Geospatial & \begin{tabular}{p{0.72\textwidth}}Fuse, overlay, register, search, analyze, annotate, and visualize   high-resolution satellite and aerial imagery, elevation data, GPS coordinates, maps, demographics, land masses, political boundaries to deliver a streaming 3D map of the entire globe \cite{popp2004countering}.\end{tabular} \\
\rowcolor[HTML]{EFEFEF} 
Grey literature & \begin{tabular}{p{0.72\textwidth}}Research that is either unpublished or has been published in   non-commercial form. Examples of grey literature include: government reports, policy statements and issues papers \cite{greyar}.\end{tabular} \\
Machine Learning & \begin{tabular}{p{0.72\textwidth}}The science of making computers learn and act like humans by feeding data  and information without begin explicitly programmed.\end{tabular} \\
\rowcolor[HTML]{EFEFEF} 
Semantic Consistency & \begin{tabular}{p{0.72\textwidth}}Exploit ontologies, taxonomies, and definitions for words, phrases, and acronyms using a variety of schemes, so users have a common and consistent understanding of the meaning of words in a specific context. Resolve semantic   heterogeneity by capitalizing on Semantic Web technologies \cite{popp2004countering}.\end{tabular} \\
White literature & \begin{tabular}{p{0.72\textwidth}}Research that has been published by an established scientific organization or group \cite{Garousi:2013:EUQ:2460999.2461003}.\end{tabular} \\
\rowcolor[HTML]{EFEFEF} 
Cyber-Physical Systems (CPS) &\begin{tabular}{p{0.72\textwidth}} These systems combine sensor networks with embedded computing to monitor and control the physical environment. Moreover, CPS are capable of making decisions and operating independently, hence these technologies can be used to monitor in real-time physical spaces remotely \cite{cpss}.\end{tabular} \\
\bottomrule
\caption{Terms used and their definitions.}
\label{tab:terms_table_pop}
\end{longtable}
\end{footnotesize}